\documentclass[
 reprint,
 superscriptaddress,
 amsmath,amssymb,
 aps,prl
]{revtex4-2}

\usepackage{MnSymbol}
\usepackage{graphicx}
\usepackage{dcolumn}
\usepackage{bm}
\usepackage{xcolor}
\usepackage{siunitx}
\usepackage{amsmath}
\usepackage[normalem]{ulem}
\usepackage{appendix}
\usepackage{placeins}

\makeatletter
\def\@caption@fignum@sep{\textbf{.}~}
\makeatother

\begin{document}

\let\oldsection\section

\title{Vapor-mediated wetting and imbibition control on micropatterned surfaces}
\author{Ze Xu}
\affiliation{Fachbereich Physik, Universität Konstanz, Konstanz, Germany}
\author{Raphael Saiseau}
\affiliation{Fachbereich Physik, Universität Konstanz, Konstanz, Germany}
\author{Olinka Ram\'irez Soto}
\affiliation{Universidad Nacional Aut\'onoma de M\'exico, M\'exico City, Mexico}
\author{Stefan Karpitschka}
\affiliation{Fachbereich Physik, Universität Konstanz, Konstanz, Germany}
\date{March 2025}

\begin{abstract} 
Wetting of micropatterned surfaces is ubiquitous in nature and key to many technological applications like spray cooling, inkjet printing, and  semiconductor processing.
Overcoming the intrinsic, chemistry- and topography-governed wetting behaviors often requires specific materials which limits applicability.
Here, we show that spreading and wicking of water droplets on hydrophilic surface patterns can be controlled by the presence of the vapor of another liquid with lower surface tension.
We show that delayed wicking arises from Marangoni forces due to vapor condensation, competing with the capillary wicking force of the surface topography.
Thereby, macroscopic droplets can be brought into an effective apparent wetting behavior, decoupled from the surface topography, but coexisting with a wicking film, cloaking the pattern.
We demonstrate how modulating the vapor concentration in space and time may guide droplets across patterns and even extract imbibed liquids, devising new strategies for coating, cleaning and drying of functional surface designs.
\end{abstract}

\maketitle

\renewcommand{\figurename}{\textbf{Fig.}}
\renewcommand{\thefigure}{\textbf{\arabic{figure}}}

Controlling the wetting behavior of droplets on surfaces is crucial to many industrial fields, such as semiconductor surface processing, inkjet printing, or coating technology~\cite{Lohse:NRP2020}.
The morphology of sessile droplets is primarily governed by surface chemistry and topography, liquid composition, and external stimuli including atmospheric composition as well as electric or magnetic fields~\cite{bonn2009wetting, gennes2004capillarity}.
Overcoming system-intrinsic wetting behaviors to, e.g., apply defect-free coatings~\cite{howison1997mathematical} or dry fragile structures non-destructively~\cite{tanaka1993mechanism} remains a ubiquitous challenge, largely because choices of surface chemistry, topography, or tolerance to electric fields are usually constrained by the final application.

Rationally designed surface topographies, often inspired by pillar or grove structures found in nature, enable unique wetting properties ranging from superhydrophilicity to superhydrophobicity~\cite{quere2008wetting}.
The intermediate regime of hemi-wicking~\cite{bico2002wetting}, where sessile droplets emit wicking films into the structures, is particularly interesting for surface processing.
Evolution and geometry of the wicking film are governed by pillar shape and pattern~\cite{kim2016dynamics,natarajan2020predicting,courbin2007imbibition}, allowing the design of dedicated functions like deposit shaping~\cite{courbin2007imbibition}, uni-directional spreading~\cite{chu2010uni, liu2017long, yang2024selective}, or liquid diodes~\cite{li2017topological}.
A complementary approach for droplet control is the spatial modulation of the liquid composition, e.g. by selective evaporation/condensation of binary~\cite{cira2015vapour,karpitschka2017marangoni} or ternary~\cite{baumgartner2022marangoni} mixtures, leading to so-called Marangoni contraction~\cite{karpitschka2017marangoni} or spreading~\cite{yang2023evaporation}:
Gradual depletion and enrichment of different chemical species toward the edge of a droplet induce surface tension gradients that drive Marangoni flows, leading to a wetting or dewetting motion, or altering the apparent contact angle~\cite{cira2015vapour,karpitschka2017marangoni,pahlavan2021evaporation,baumgartner2022marangoni,yang2023evaporation}.
Counteracting topography-induced hemi-wicking with vapor-mediated Marangoni forces could leverage dynamic control over wetting and imbibition on textured surfaces.
Crucially, this approach could also tolerate constraints on topography and/or material choice imposed by the final application.
Yet, this combination remains unexplored in scientific literature.

\begin{figure*}[!ht]
\centering
\includegraphics{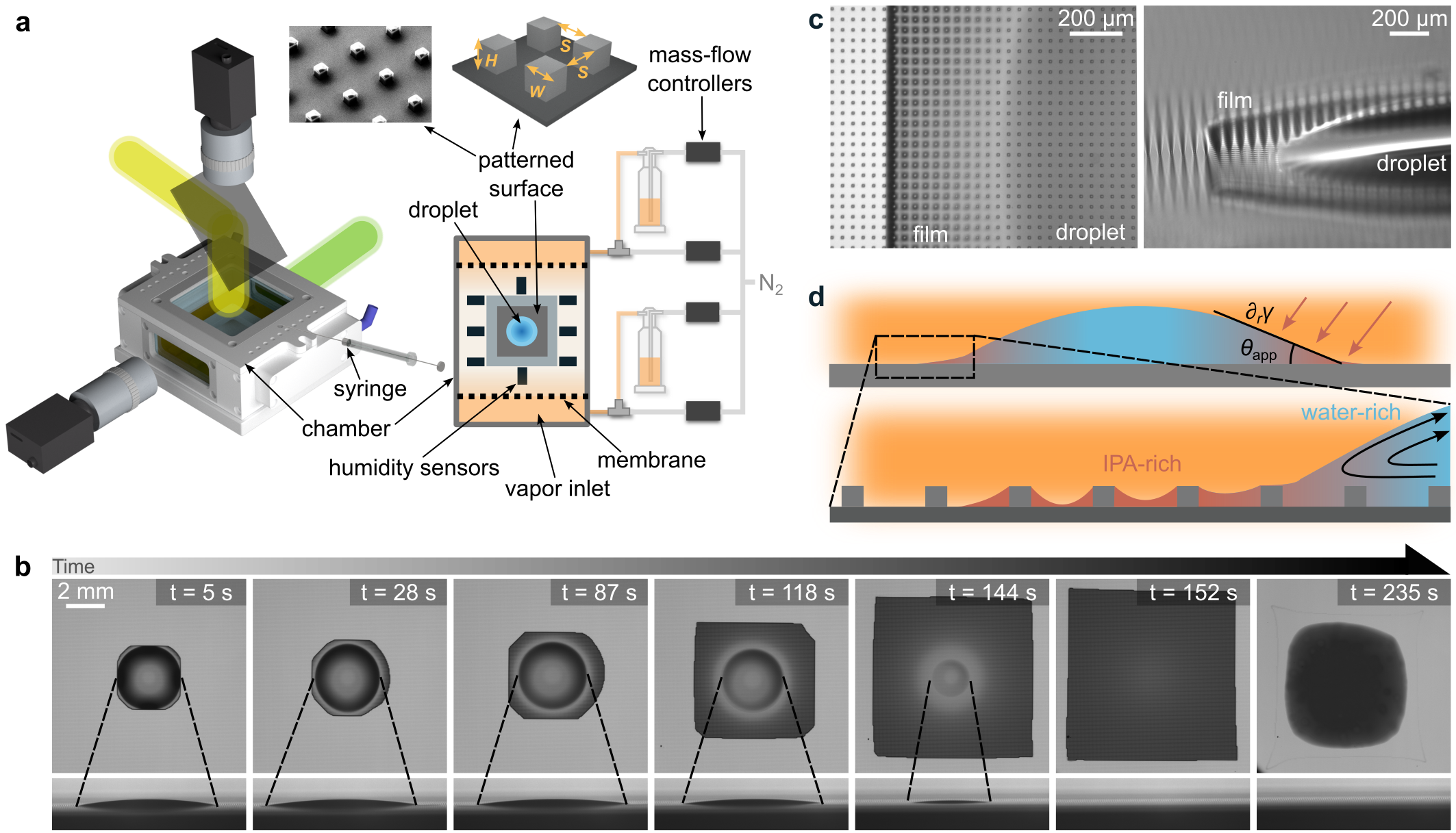}
\caption{\textbf{Vapor-mediated wetting control on micro-patterned surfaces.} \textbf{a},~Schematic of the experimental setup and SEM image of a micropillar decorated silicon wafer ($S=\SI{30}{\micro\meter}$, $W=H=\SI{10}{\micro\meter}$). \textbf{b},~Wicking of water into hydrophilic textures is delayed more than 100-fold in an isopropyl alcohol (IPA) vapor atmosphere (images from top and side aspects, time $t$ since drop deposition). A Marangoni-contracted droplet coexists with a wicking film around it which effectively cloaks the pattern for the droplet. \textbf{c},~High magnification photographs of the edge region of the drop-on-film configuration. \textbf{d},~Schematic of the liquid configuration: a macroscopic droplet, connected to a imbibed film that thins and enriches in IPA away from the droplet.}
\label{fig:experiment}
\end{figure*}

Here we demonstrate that the spreading and imbibition dynamics of sessile droplets of water/isopropyl alcohol (IPA) mixtures on hemi-wicking surface topographies are upended by selective condensation/evaporation in a non-equilibrium vapor environment, although both liquids wet and wick the patterns.
The surface structures enhance wettability because the equilibrium (microscopic) contact angle is below a critical value governed by the roughness parameters of the texture, thus generating a capillary wicking force~\cite{bico2002wetting}.
Without compositional gradients, the resulting dynamics are akin to the Washburn law, describing the liquid front wicking into a porous material~\cite{ishino2007wicking, gambaryan2014liquids}.
In the presence of IPA vapor, however, we observe an extreme delay or, initially, even complete inhibition of the wicking, leading to a prolonged coexistence of a macroscopic droplet body surrounded by an imbibition film of finite extent.
Remarkably, the droplet remains free to move as the film cloaks the surface topography.
We demonstrate that this phenomenon is driven by surface tension gradients, arising from the absorption of vapor of a lower surface tension liquid.
The resulting Marangoni flows compensate for the capillary action of the micro-pattern, normally responsible for spreading and wicking.
Thus, the phenomenon is not specific to water and IPA, but generalizes to miscible pairs where the condensing component has a lower surface tension than the liquid.
Exploiting this antagonism, we explore strategies for droplet manipulation and even extraction of imbibed liquids.
The dual nature of the phenomenon induces a significant degree of freedom to the choices of surface topography and liquid parameters, thus holding a promising potential for technological applications.

\section{Results}

\subsection{Experimental Approach}

Experiments were performed on canonical surface topographies, i.e. micropillar arrays fabricated by direct laser writing of an SU-8 photoresist on silicon wafers (Fig.~\ref{fig:experiment}\textbf{a}). 
Unless stated otherwise, pillars were cube-shaped with $\SI{10}{\micro\meter}$ edge length, arranged on a square lattice with varying spacing $S$, and hydrophilized by plasma treatment before use. 
Pure water droplets in dry or moist atmospheres are fully wicked into these patterns within less than one second (Extended Data Fig.~\ref{SIfig:immediate} and Supplementary Video 7).
The same happens for pure isoprolyl alcohol (IPA) droplets in dry air or IPA vapor.
Experiments were performed in an environmental chamber (Fig. ~\ref{fig:experiment}\textbf{a}) at constant room temperature ($\sim\SI{21}{\celsius}$), constantly feeding nitrogen gas with the desired vapor content to two inlet chambers which, across membranes, diffusively exchange gases with the main chamber (Fig.~\ref{fig:experiment}\textbf{a}).
Vapor compositions were varied by bubbling nitrogen through gas wash bottles containing mixtures of IPA (mass fraction $\tilde{c}$) and water.
The wetting and wicking dynamics were imaged from top and side through windows.
Further details are provided in the Online Methods and the Supplementary Information.

\begin{figure*}[!ht]
\centering
\includegraphics[scale=1]{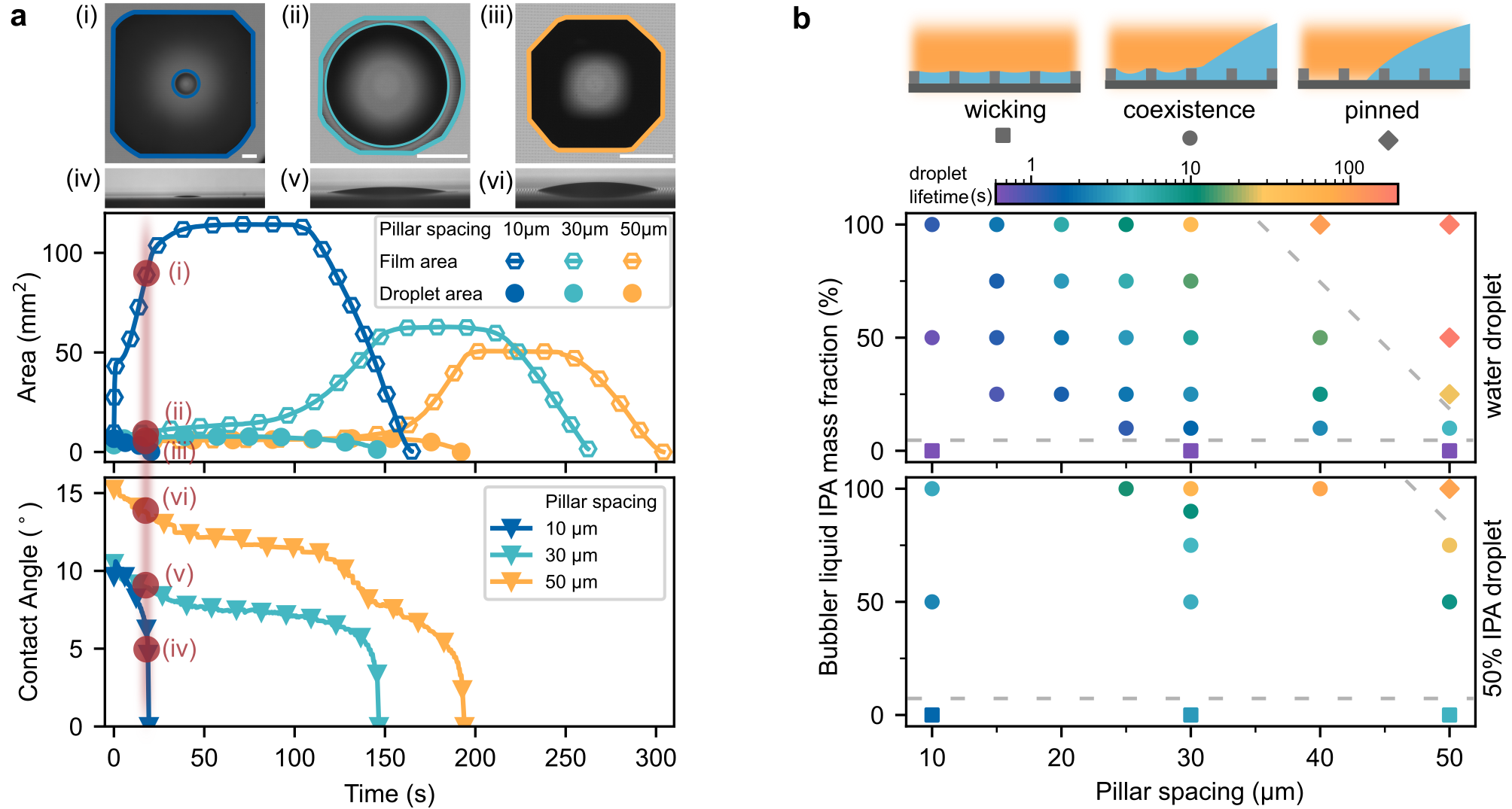}
\caption{\textbf{Drop-film coexistence with varying texture spacing, vapor and droplet compositions.} \textbf{a},~Evolution of drop and film footprint area (top) and apparent contact angle $\theta_{\mathrm{app}}$ of the drop (bottom) on surfaces with varying pillar spacing $S$ in IPA vapor environment. Images show top and side aspects at $t=\SI{18}{\second}$ (scale bars indicate $\SI{1}{\milli\meter}$). Smaller pillar spacings enhance the wicking force, leading to shorter drop and film lifetimes and smaller apparent contact angles of the drop. For $S=\SI{50}{\micro\meter}$, wicking is completely prevented initially (images (iii) and (vi)). \textbf{b},~Phase diagram of wetting behaviors \emph{vs} pillar spacing and IPA vapor concentration for water droplets (top) and 50\% IPA aqueous droplets (bottom). Color indicates the lifetime of the contracted droplet (cf. Extended Data Fig.~\ref{SIfig:Drop_lifetime}). The IPA vapor concentration is characterized by the mass fraction of liquid IPA in the bubbler, $\tilde{c}$, indicative of the mismatch of drop-vapor equilibrium  (see Supplementary Material).}
\label{fig:phasediagram}
\end{figure*}

\subsection{Delayed Wicking}

Depositing a droplet of pure water at time $t=0$ onto a surface with pillar spacing $S=\SI{30}{\micro\meter}$ in an IPA-vapor environment (Fig.~\ref{fig:experiment}\textbf{b} and Supplementary Video 1), the droplet does not readily spread across or wick into the surface pattern, as was observed in dry or humid ambient (Extended Data Fig.~\ref{SIfig:immediate}).
Rather, the droplet remains in a spherical-cap-like shape, while a wicking film slowly emerges from its periphery ($\gtrsim 100$ times slower than in the pure case).
Long-distance video microscopy from the top and the side reveals the local topography of the liquid surface (Fig.~\ref{fig:experiment}\textbf{c}), which is sketched in Fig.~\ref{fig:experiment}\textbf{d}:
The droplet surface merges onto a pillar-imbibed film over a narrow transition zone, with an apparent contact angle $\theta_{\text{app}}\sim\SI{11}{\degree}$.
Sloped liquid surfaces refract light from the episcopic illumination, appearing darker in the top view image.
Close to the droplet, the film surface appears evenly bright, indicating a flat film, level to the pillar tops.
Approaching the outer boundary of the film,
pillars increasingly pinch on the thinning film, leading to dark halos around them.
The film terminates in a steep (dark) meniscus just before the first dry row of pillars.
Over time, the film expands in a zipping-like motion~\cite{courbin2007imbibition} (Extended Data Fig.~\ref{SIfig:zipping}):
Once the terminal meniscus touches a pillar of the next row, capillary forces ``pull along'' the film to adjacent pillars.
The droplet, in contrast, does not pin to the pillars but assumes a circular footprint and remains highly mobile, reminiscent of Marangoni-contracted droplets on flat surfaces~\cite{tsoumpas2015effect,cira2015vapour,benusiglio2018two,karpitschka2017marangoni,Ramirez:PRF2022,charlier2022water}.

\begin{figure*}[!ht]
\centering
\includegraphics[scale=1]{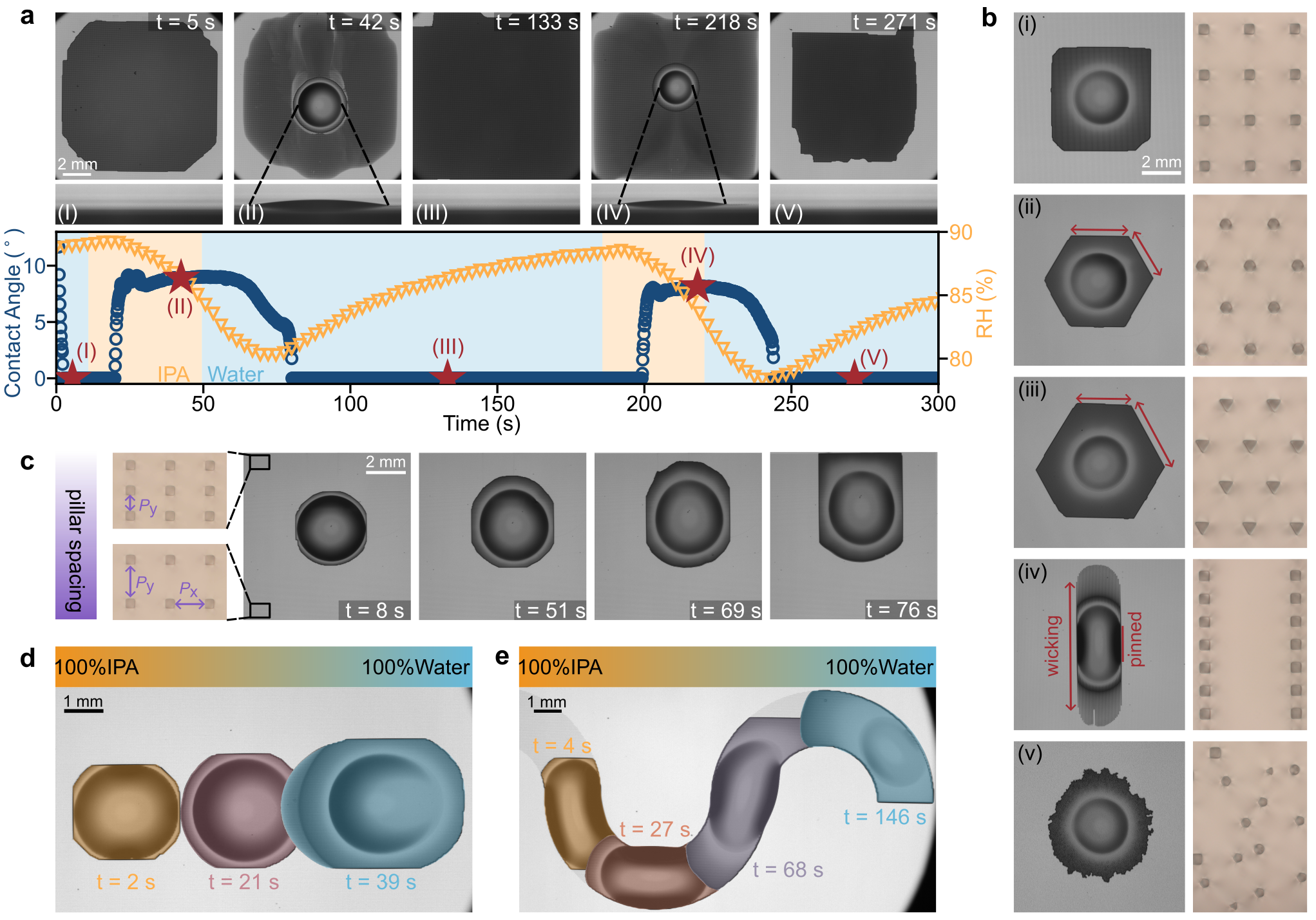}
\caption{\textbf{Vapor-mediated droplet manipulation on patterned surfaces.}
\textbf{a},~Repeatable liquid extraction from fully imbibed films by switching between water and IPA vapor in the atmosphere. We plot the apparent contact angle of the droplet, $\theta_{\text{app}}$ (blue), and the relative humidity in the chamber, $RH$ (yellow), \emph{vs} time $t$ since deposition (red stars indicate picture times, background color indicates mass flow controller setting. $S=\SI{30}{\micro\meter}$, $W=H=\SI{10}{\micro\meter}$).
\textbf{b},~Impact of lattice symmetry and pillar shape, for similar pillar area fraction $\phi_{p}\sim 0.065\pm 0.003$ (pure water drop in IPA vapor):
(i)~square pillars on a square lattice, 
(ii)~hexagonal pillars on a hexagonal lattice, 
(iii)~triangular pillars on a hexagonal lattice, 
(iv)~square pillars on a rectangular lattice, 
and (v)~random shapes in random arrangement. 
Wicking forces are governed by the lattice, but pinning is modified by pillar shape.
\textbf{c},~Directional film spreading and drop motion on a square lattice with spacing gradient ($S_y = \SIrange{35}{15}{\micro\meter}$, $S_x = \SI{30}{\micro\meter}$, pure water drop in IPA vapor).
\textbf{d},~Droplet motion on a homogeneous square lattice ($S = \SI{30}{\micro\meter}$) with IPA vapor gradient.
\textbf{e},~Droplet forced along a square pillar lattice within ''S''-shaped bounds ($S = \SI{30}{\micro\meter}$) by a vapor gradient in $x$-direction.}
\label{fig:manipulation}
\end{figure*}

\subsection{Interplay of Wicking and Contraction}

We quantify the competition between capillary wicking and Marangoni contraction by varying pillar spacing, IPA vapor concentration (characterized by the liquid IPA mass fraction in the bubbler, $\tilde{c}$), and droplet composition: narrower pillar spacings $S$ are expected to generate a stronger wicking force (capillary pressure $p\sim S^{-1}$), while adding liquid IPA to the droplet or reducing the partial pressure of IPA in the atmosphere would reduce the rate of condensation and thus attenuate the Marangoni force.
We start by varying $S$, keeping $\tilde{c}=1$ and pure water as initial drop composition.
The droplet and film dynamics can readily be quantified by the evolution of their footprint areas.
Additionally, the apparent contact angle of the droplet indicates the strength of the contracting Marangoni force. We plot these quantities in Fig.~\ref{fig:phasediagram}\textbf{a} as a function of time.
For $S=\SI{30}{\micro\meter}$ (the case shown in Fig.~\ref{fig:experiment}), the plot indicates coexistence of drop and wicking film for about $\sim\SI{150}{\second}$, during which the apparent contact angle slowly decreases while the film and droplet areas remain approximately constant.
This stage of delayed wicking and drop/film coexistence is significantly diminished when decreasing the pillar spacing to $S=\SI{10}{\micro\meter}$ (Extended Data Fig.~\ref{SIfig:delayedP10P50}\textbf{a} and Supplementary Video 2), yet still about an order of magnitude larger than the pure case (Extended Data Fig.~\ref{SIfig:immediate}).
When increasing the pillar spacing to $S=\SI{50}{\micro\meter}$ (Extended Data Fig.~\ref{SIfig:delayedP10P50}\textbf{b} and Supplementary Video 2), wicking is not only delayed, but completely inhibited for $\sim\SI{128}{\second}$.
During this stage, the droplet is pinned to columns, rows, or diagonals of pillars, giving it an octagonal footprint shape.

High resolution imaging of the contact line zone reveals the mechanism (Extended data Fig.~\ref{SIfig:depinning} and Supplementary Video 8):
The apparent contact angle of the contracted droplet remains larger than the aspect ratio of the gap between pillars, $\tan\theta_{\text{app}}>H/S$.
Because the free surface is pinned to the caps of the last wetted row of pillars, the liquid now cannot reach the next row and the wicking force cannot develop, analogous to static wetting~\cite{courbin2007imbibition}.
This geometric threshold is different from the energetically motivated transition from partial wetting to hemi-wicking~\cite{bico2002wetting,quere2008wetting} because the apparent angle is determined by non-equilibrium processes while the microscopic angle is always below the critical angle for hemi-wicking~\cite{karpitschka2017marangoni}.
As IPA continues being absorbed into the droplet, contraction weakens until the wicking condition is reached, starting along the diagonal where row spacings are smallest~\cite{courbin2007imbibition}.
About $\SI{10}{\second}$ later, this condition is also met along the lattice axes, and a circular, Marangoni-contracted droplet emerges on top of an imbibed film, impassive of the substrate pattern and free to move.

Toward the end of the coexistence phase (Figs.~\ref{fig:phasediagram}\textbf{a} and~\ref{fig:experiment}\textbf{b}, $t\gtrsim\SI{118}{\second}$), the wicking film grows, drawing volume from the droplet. Remarkably, the growth rate always follows a standard hemi-wicking power-law \cite{natarajan2020predicting,ishino2007wicking,srivastava2010unified,kim2011hydrodynamics,kim2016dynamics,krishnan2019simple}, but the prefactor is decreased by about an order of magnitude as compared to the case without Marangoni flows (see Extended Data Fig.~\ref{SIfig:Film_Wicking_experiments}).
Once the droplet merged into the imbibed film, expansion ceases and the final evaporation regime begins (Fig.~\ref{fig:experiment}\textbf{b}, $t\gtrsim\SI{152}{\second}$).

In Fig.~\ref{fig:phasediagram}\textbf{b}, we provide a phase diagram that summarizes these three distinct behaviors (immediate wicking $\square$, coexistence $\circ$, pinning $\Diamond$) as a function of $S$ and $\tilde{c}$, for initially pure water droplets (top) and droplets with 50 wt\% IPA (bottom).
The lifetime of the contracted droplet is indicated as color code.
Generally, droplet lifetime increases with $S$ and $\tilde{c}$ because wicking forces decrease with pillar spacing while Marangoni forces are sustained longer in higher IPA vapor concentrations (see Extended Data Figs.~\ref{SIfig:evolP30}-\ref{SIfig:flat&p50} for plots of the footprint areas over time (\ref{SIfig:evolP30}), drop lifetimes over $S$ (\ref{SIfig:Drop_lifetime}), and apparent contact angles and drop lifetimes for different drop compositions on flat and textured surfaces (\ref{SIfig:flat&p50})).
Already small amounts of IPA vapor delay the wicking process noticeably, especially for large pillar spacings, so that ``immediate wicking'' is limited to pure cases.
Since the pinned state emerges when the apparent contact angle from Marangoni contraction exceeds the pattern aspect ratio, that regime depends on both vapor composition and pattern spacing.

\section{Discussion}

The prolonged lifetime of contracted droplets on top of an imbibed film allows for unprecedented liquid manipulation strategies on textured surfaces.
A collection of possible manipulation sequences is shown in Fig.~\ref{fig:manipulation}.
We start by showing that fully imbibed liquids can reversibly be extracted into contracted droplets (Fig.~\ref{fig:manipulation}\textbf{a} and Supplementary Video 3):
A water droplet deposited in a moist atmosphere is immediately fully wicked into the texture.
By use of the gas mixing system, we then switch from pure water vapor to pure IPA vapor.
This causes an instability of the film which develops into a Marangoni-contracted droplet ($\theta_{\text{app}}\sim\SI{9}{\degree}$).
The droplet is still surrounded by an extended film that is, on average, much thinner than the pillar height, rendered dark by capillary rise on the pillar side walls.
Returning to pure water vapor in the atmosphere, the droplet is imbibed back into the surface texture, and the extraction process can be repeated by switching between vapor types.

Different film shapes (e.g. square, hexagon, elongated hexagon, irregular) emerge from the interplay of pillar shapes and lattice symmetries, as shown in Fig. ~\ref{fig:manipulation}\textbf{b} and Supplementary Video 4, caused by anisotropic spreading resistances, analogous to the shape formation on wicking~\cite{courbin2007imbibition, courbin2009dynamics, sontheimer2024numerical} or non-wicking patterned surfaces~\cite{lou2022polygonal, raj2014high, al2023binary}.
The pillar shape modulates the film wicking, as highlighted in panels (ii) and (iii): the lattice is identical in both cases, but the pronounced edges of the triangular pillars in (iii) increase pinning forces in some directions, breaking the equilateral symmetry of the final hexagonal film shape.

For strong lattice anisotropies, wicking and pinning states may coexist along different axes (Fig. ~\ref{fig:manipulation}\textbf{b}(iv)), leading also to highly anisotropic droplet mobility.
In the case shown in Fig. ~\ref{fig:manipulation}\textbf{b}(iv), $S_x = \SI{70}{\micro\meter}$ and $S_y = \SI{10}{\micro\meter}$ effectively provides rails keeping the droplet mobile only in $y$-direction.
The drop/film coexistence effect pertains to random arrangements (panel (v)), where a circular, contracted droplet is surrounded by an irregular film, demonstrating that our strategy may be applicable to wide variety or surface patterns that need not be specifically designed for the dewetting process.

Based on the observation that smaller pillar spacings give rise to stronger wicking forces, surfaces can also readily be designed to induce droplet motion and thus directional wetting/drying:
Fig.~\ref{fig:manipulation}\textbf{c} (Supplementary Video 5) shows a contracted droplet (pure water in IPA vapor) on a lattice with a unidirectional spacing gradient, $S_y = \SIrange{35}{15}{\micro\meter}$, while $S_x = \SI{30}{\micro\meter}$.
After depositing the droplet, the liquid wicks faster in the direction of decreasing pillar spacing.
Because droplet motion and film spreading are coupled, the entire liquid body moves down the pillar spacing gradient.

The sensitivity to vapor composition also allows for contactless droplet manipulation on micropatterned substrates, which is demonstrated in Fig.~\ref{fig:manipulation}\textbf{d} (Supplementary Video 6).
There, IPA vapor was injected on the left, water on the right end of the environmental chamber, generating a constant and homogeneous vapor composition gradient.
In this environment, the water droplet moves toward the water vapor side, caused by faster IPA absorption and thus stronger Marangoni forces on the opposite side, analogous to the case on flat surfaces~\cite{cira2015vapour}.
Utilizing both, preferential spreading and vapor gradient actuation, droplets can be guided along complex paths, as shown in Fig.~\ref{fig:manipulation}\textbf{e} and Supplementary Video 6.
Here, a pillar pattern ($S=\SI{30}{\micro\meter}$) bounded by an S-shaped outline guides the liquid, and a vapor gradient analogous to panel \textbf{d} induces the motion.

The examples from Fig.~\ref{fig:manipulation} demonstrate the versatility of manipulating droplets on textured surfaces dynamically through vapors in the environment.
The effect is easily implemented, highly reproducible, and robust against variations of pattern properties and vapor compositions  (Fig.~\ref{fig:phasediagram}).
It also generalizes to arbitrary combinations of liquids, requiring only miscibility and a lower surface tension of the excess vapor component.
In contrast to lubricant infused surfaces~\cite{Wong:N2011}, here the liquid components are volatile and miscible, thus eliminating the need for separate infusion and removal of lubricant, akin to self-lubricating droplets~\cite{tan2023self}.
Notably, the wicking force is governed only by ratios of the geometric parameters $S$, $W$ and $H$, ensuring broad applicability across a wide range of scales.
Still, viscous dissipation and thus the rate of wicking depends on the scale of the texture, possibly expanding the coexistence regime for smaller structures (see SI section \emph{Wicking dynamics} and~\cite{bico2002wetting}).
This inherent scalability thus leaves a significant degree of freedom for the design of surface patterns or choice of chemicals, facilitating its application in the processing of surface microstructures, liquid collection and extraction techniques, or sensing device applications.

\let\section\oldsection

\section*{Online Methods}
\subsection{Preparation of surface patterns}
The micropillar arrays were fabricated by direct laser writing lithography of a negative photoresist.
First, the silicon wafers  (CZ-Si, (100), $\SI{500 \pm 50}{\micro\meter}$, p-type, MicroChemicals GmbH) were cleaved into square pieces of $\qtyproduct{25 x 25}{\milli\meter}$. 
The cleaved substrates were then sequentially rinsed with deionized water, acetone, and isopropanol, followed by drying with nitrogen gas and baking at 150 $^\circ C$ to remove moisture thoroughly. 
Next, the photoresist (SU-8 3010, KAYAKU Advanced Materials) was spin-coated onto the substrate (initial spin at 500~rpm for 30~s with an acceleration rate of 100~rpm/s, followed by a second spin at 3000~rpm for 50~s with an acceleration of 300~rpm/s). 
Subsequently, the wafer was soft-baked at $\SI{95}{\celsius}$
for 5 minutes. 
Then, the wafer was exposed in the direct laser writer (POLOS NanoWriter Advanced, PM-100). 
Thereafter, a post-exposure baking was conducted at $\SI{95}{\celsius}$
for 3 minutes.
Finally, the patterned substrate was rinsed thoroughly in SU-8 developer and dry-blown with nitrogen.

\subsection{Experimental setup}
The experimental setup consisted of a custom-designed environmental chamber (Internal dimensions $L=\SI{85}{\milli\meter}$ x  $W=\SI{75}{\milli\meter}$ x $H=\SI{33}{\milli\meter}$), a vapor generation system, and an imaging system that recorded top and side aspects of the droplets through windows in the chamber.
The chamber was well-sealed against the ambient to ensure stable vapor environments inside.
The left and right walls of the main chamber were comprised of cloth membranes, behind which two separate spout chambers were located.
These spout chambers were flushed with a continuous stream of nitrogen containing the vapors.
This way, vapors were only diffusively injected into the main chamber, generating a stable, quiescent atmosphere.
Eight humidity sensors (Honeywell HIH-4000) were integrated into the main chamber to measure the humidity distribution.

The vapors were prepared by a custom-built gas mixing system, consisting of two identical lines, one for each side of the chamber.
Each line consisted of two mass flow controllers (MKS Instruments, GE50A, 500 sccm), where the output flux of one controller was bubbled through a gas wash bottle, containing water or IPA or a mixture of them, depending on the desired vapor composition.
We used ultra pure water (Milli-Q, resistivity $\SI{18.2}{\mega\ohm\centi\meter}$) and isopropanol (VWR, 99.7+\%) as liquids for droplets and in the gas wash bottles.
Vapor compositions are quantified by the mass fraction of the liquids in the bubblers (see Extended Data Figure~\ref{SIfig:Activity_RH_IPA-water} for the relation to actual vapor partial pressures).
The gas wash bottles were placed into a thermal bath ($T=\SI{22}{\celsius}$) to compensate for evaporative cooling.
The two fluxes were then unified and injected into the spout chamber.
The total mass flux (i.e., the sum of the two fluxes) was kept constant across all experiments, only the ratio of the two fluxes was varied.
Experiments with static, homogeneous atmosphere were performed with 100\% of the flux passing through the gas wash bottle.
Note that the vapor always remained undersaturated to prevent condensation on solid surfaces (see Extended Data Figure~\ref{SIfig:Activity_RH_IPA-water}) and keep the drying times of the films finite.
Liquids in the bottles were replaced after each experiment to ensure constant composition of the liquids in the bubblers.
The vapor composition gradient was created by filling pure isopropanol and pure water to the gas wash bottles for the two sides.
For the extraction experiments where we switched between different vapors, gas lines were reconfigured to simultaneously switch vapors on both sides.

The imaging system consisted of two collimated illuminators and two cameras.
For the top view perspective, the light of a custom-designed Koehler-type illuminator was projected by a 50/50 beamsplitter onto the sample, reflected by the substrate, and recorded by a macro lens (Schneider optics \SI{50}{mm}) and a monochromatic CMOS camera (FLIR Grasshopper).
High magnification images were taken with a zoom lens (Thorlabs MVL12X3Z).
The side aspect was recorded with a telecentric lens (Thorlabs 1.0X Bi-Telecentric lens) and an identical CMOS camera.
Illumination was provided from the rear with a telecentric illuminator (Edmund optics, $\SI{52}{\milli\meter}$).
Image analysis was done by a self-developed Python script.

\subsection{Experimental procedure}
Before each experiment, the substrate was cleaned with deionized water and isopropanol, then dried using nitrogen gas. 
Then, the substrate was plasma treated in a plasma oven ($\SI{13.56}{\mega\hertz}$, Diener electronic GmbH) for one minute and immediately transferred into the chamber.
After the vapor environment in the chamber stabilized according to the readings from the humidity sensors, liquid droplets with volumes from $\SI{0.5}{\micro\litre}$ to $\SI{1}{\micro\litre}$ were deposited onto the substrate with a clean precision syringe (Hamilton Gas-tight Series 1700, $\SI{10}{\micro\litre}$)). 
The delay between plasma treatment and droplet deposition was always less than 3 minutes.
To minimize exchange between the atmosphere in the chamber and the liquid in the syringe prior to depositing the droplet, air was drawn into the needle before inserting it into the chamber through a small hole next to the top window.

\subsection{Acknowledgements}
We thank Nate Cira, Adrien Benusiglio, Manu Prakash and Hans Riegler for discussions. We are indebted to Louis Kukk, Kris Handtke and Wolf Keiderling for assistance with the technical realization of the setup. 
We acknowledge the use of the experimental equipment and the expert support concerning its usage and data analysis provided by the Nanostructure Laboratory at the University of Konstanz.
R.S. was supported by the German Research Foundation (DFG, grant no. 422877263). Z.X. was supported by China Scholarship Council (CSC202206090036).

\subsection{Author contributions}
S.K. conceived the study, Z.X., O.R. and S.K. designed the experiments, Z.X. performed the experiments, Z.X. and R.S. analyzed data. All authors contributed to the interpretation of the data, discussions of the results, and writing of the manuscript.

\subsection{Competing interests}
The authors declare no competing interests.


\begin{thebibliography}{40}%
\makeatletter
\providecommand \@ifxundefined [1]{%
 \@ifx{#1\undefined}
}%
\providecommand \@ifnum [1]{%
 \ifnum #1\expandafter \@firstoftwo
 \else \expandafter \@secondoftwo
 \fi
}%
\providecommand \@ifx [1]{%
 \ifx #1\expandafter \@firstoftwo
 \else \expandafter \@secondoftwo
 \fi
}%
\providecommand \natexlab [1]{#1}%
\providecommand \enquote  [1]{``#1''}%
\providecommand \bibnamefont  [1]{#1}%
\providecommand \bibfnamefont [1]{#1}%
\providecommand \citenamefont [1]{#1}%
\providecommand \href@noop [0]{\@secondoftwo}%
\providecommand \href [0]{\begingroup \@sanitize@url \@href}%
\providecommand \@href[1]{\@@startlink{#1}\@@href}%
\providecommand \@@href[1]{\endgroup#1\@@endlink}%
\providecommand \@sanitize@url [0]{\catcode `\\12\catcode `\$12\catcode
  `\&12\catcode `\#12\catcode `\^12\catcode `\_12\catcode `\%12\relax}%
\providecommand \@@startlink[1]{}%
\providecommand \@@endlink[0]{}%
\providecommand \url  [0]{\begingroup\@sanitize@url \@url }%
\providecommand \@url [1]{\endgroup\@href {#1}{\urlprefix }}%
\providecommand \urlprefix  [0]{URL }%
\providecommand \Eprint [0]{\href }%
\providecommand \doibase [0]{https://doi.org/}%
\providecommand \selectlanguage [0]{\@gobble}%
\providecommand \bibinfo  [0]{\@secondoftwo}%
\providecommand \bibfield  [0]{\@secondoftwo}%
\providecommand \translation [1]{[#1]}%
\providecommand \BibitemOpen [0]{}%
\providecommand \bibitemStop [0]{}%
\providecommand \bibitemNoStop [0]{.\EOS\space}%
\providecommand \EOS [0]{\spacefactor3000\relax}%
\providecommand \BibitemShut  [1]{\csname bibitem#1\endcsname}%
\let\auto@bib@innerbib\@empty
\bibitem [{\citenamefont {Lohse}\ and\ \citenamefont
  {Zhang}(2020)}]{Lohse:NRP2020}%
  \BibitemOpen
  \bibfield  {author} {\bibinfo {author} {\bibfnamefont {D.}~\bibnamefont
  {Lohse}}\ and\ \bibinfo {author} {\bibfnamefont {X.}~\bibnamefont {Zhang}},\
  }\bibfield  {title} {\bibinfo {title} {Physicochemical hydrodynamics of
  droplets out of equilibrium},\ }\href
  {https://doi.org/10.1038/s42254-020-0199-z} {\bibfield  {journal} {\bibinfo
  {journal} {Nature Reviews Physics}\ }\textbf {\bibinfo {volume} {2}},\
  \bibinfo {pages} {426} (\bibinfo {year} {2020})}\BibitemShut {NoStop}%
\bibitem [{\citenamefont {Bonn}\ \emph {et~al.}(2009)\citenamefont {Bonn},
  \citenamefont {Eggers}, \citenamefont {Indekeu}, \citenamefont {Meunier},\
  and\ \citenamefont {Rolley}}]{bonn2009wetting}%
  \BibitemOpen
  \bibfield  {author} {\bibinfo {author} {\bibfnamefont {D.}~\bibnamefont
  {Bonn}}, \bibinfo {author} {\bibfnamefont {J.}~\bibnamefont {Eggers}},
  \bibinfo {author} {\bibfnamefont {J.}~\bibnamefont {Indekeu}}, \bibinfo
  {author} {\bibfnamefont {J.}~\bibnamefont {Meunier}},\ and\ \bibinfo {author}
  {\bibfnamefont {E.}~\bibnamefont {Rolley}},\ }\bibfield  {title} {\bibinfo
  {title} {Wetting and spreading},\ }\href@noop {} {\bibfield  {journal}
  {\bibinfo  {journal} {Reviews of modern physics}\ }\textbf {\bibinfo {volume}
  {81}},\ \bibinfo {pages} {739} (\bibinfo {year} {2009})}\BibitemShut
  {NoStop}%
\bibitem [{\citenamefont {de~Gennes}\ \emph {et~al.}(2004)\citenamefont
  {de~Gennes}, \citenamefont {Brochard-Wyart},\ and\ \citenamefont
  {Qu{\'e}r{\'e}}}]{gennes2004capillarity}%
  \BibitemOpen
  \bibfield  {author} {\bibinfo {author} {\bibfnamefont {P.-G.}\ \bibnamefont
  {de~Gennes}}, \bibinfo {author} {\bibfnamefont {F.}~\bibnamefont
  {Brochard-Wyart}},\ and\ \bibinfo {author} {\bibfnamefont {D.}~\bibnamefont
  {Qu{\'e}r{\'e}}},\ }\href@noop {} {\emph {\bibinfo {title} {Capillarity and
  wetting phenomena: drops, bubbles, pearls, waves}}}\ (\bibinfo  {publisher}
  {Springer},\ \bibinfo {year} {2004})\BibitemShut {NoStop}%
\bibitem [{\citenamefont {Howison}\ \emph {et~al.}(1997)\citenamefont
  {Howison}, \citenamefont {Moriarty}, \citenamefont {Ockendon}, \citenamefont
  {Terrill},\ and\ \citenamefont {Wilson}}]{howison1997mathematical}%
  \BibitemOpen
  \bibfield  {author} {\bibinfo {author} {\bibfnamefont {S.}~\bibnamefont
  {Howison}}, \bibinfo {author} {\bibfnamefont {J.}~\bibnamefont {Moriarty}},
  \bibinfo {author} {\bibfnamefont {J.}~\bibnamefont {Ockendon}}, \bibinfo
  {author} {\bibfnamefont {E.}~\bibnamefont {Terrill}},\ and\ \bibinfo {author}
  {\bibfnamefont {S.}~\bibnamefont {Wilson}},\ }\bibfield  {title} {\bibinfo
  {title} {A mathematical model for drying paint layers},\ }\href@noop {}
  {\bibfield  {journal} {\bibinfo  {journal} {Journal of Engineering
  Mathematics}\ }\textbf {\bibinfo {volume} {32}},\ \bibinfo {pages} {377}
  (\bibinfo {year} {1997})}\BibitemShut {NoStop}%
\bibitem [{\citenamefont {Tanaka}\ \emph {et~al.}(1993)\citenamefont {Tanaka},
  \citenamefont {Morigami},\ and\ \citenamefont {Atoda}}]{tanaka1993mechanism}%
  \BibitemOpen
  \bibfield  {author} {\bibinfo {author} {\bibfnamefont {T.}~\bibnamefont
  {Tanaka}}, \bibinfo {author} {\bibfnamefont {M.~M.~M.}\ \bibnamefont
  {Morigami}},\ and\ \bibinfo {author} {\bibfnamefont {N.~A.~N.}\ \bibnamefont
  {Atoda}},\ }\bibfield  {title} {\bibinfo {title} {Mechanism of resist pattern
  collapse during development process},\ }\href@noop {} {\bibfield  {journal}
  {\bibinfo  {journal} {Japanese journal of applied physics}\ }\textbf
  {\bibinfo {volume} {32}},\ \bibinfo {pages} {6059} (\bibinfo {year}
  {1993})}\BibitemShut {NoStop}%
\bibitem [{\citenamefont {Qu{\'e}r{\'e}}(2008)}]{quere2008wetting}%
  \BibitemOpen
  \bibfield  {author} {\bibinfo {author} {\bibfnamefont {D.}~\bibnamefont
  {Qu{\'e}r{\'e}}},\ }\bibfield  {title} {\bibinfo {title} {Wetting and
  roughness},\ }\href@noop {} {\bibfield  {journal} {\bibinfo  {journal} {Annu.
  Rev. Mater. Res.}\ }\textbf {\bibinfo {volume} {38}},\ \bibinfo {pages} {71}
  (\bibinfo {year} {2008})}\BibitemShut {NoStop}%
\bibitem [{\citenamefont {Bico}\ \emph {et~al.}(2002)\citenamefont {Bico},
  \citenamefont {Thiele},\ and\ \citenamefont
  {Qu{\'e}r{\'e}}}]{bico2002wetting}%
  \BibitemOpen
  \bibfield  {author} {\bibinfo {author} {\bibfnamefont {J.}~\bibnamefont
  {Bico}}, \bibinfo {author} {\bibfnamefont {U.}~\bibnamefont {Thiele}},\ and\
  \bibinfo {author} {\bibfnamefont {D.}~\bibnamefont {Qu{\'e}r{\'e}}},\
  }\bibfield  {title} {\bibinfo {title} {Wetting of textured surfaces},\
  }\href@noop {} {\bibfield  {journal} {\bibinfo  {journal} {Colloids and
  Surfaces A: Physicochemical and Engineering Aspects}\ }\textbf {\bibinfo
  {volume} {206}},\ \bibinfo {pages} {41} (\bibinfo {year} {2002})}\BibitemShut
  {NoStop}%
\bibitem [{\citenamefont {Kim}\ \emph {et~al.}(2016)\citenamefont {Kim},
  \citenamefont {Moon},\ and\ \citenamefont {Kim}}]{kim2016dynamics}%
  \BibitemOpen
  \bibfield  {author} {\bibinfo {author} {\bibfnamefont {J.}~\bibnamefont
  {Kim}}, \bibinfo {author} {\bibfnamefont {M.-W.}\ \bibnamefont {Moon}},\ and\
  \bibinfo {author} {\bibfnamefont {H.-Y.}\ \bibnamefont {Kim}},\ }\bibfield
  {title} {\bibinfo {title} {Dynamics of hemiwicking},\ }\href@noop {}
  {\bibfield  {journal} {\bibinfo  {journal} {Journal of Fluid Mechanics}\
  }\textbf {\bibinfo {volume} {800}},\ \bibinfo {pages} {57} (\bibinfo {year}
  {2016})}\BibitemShut {NoStop}%
\bibitem [{\citenamefont {Natarajan}\ \emph {et~al.}(2020)\citenamefont
  {Natarajan}, \citenamefont {Jaishankar}, \citenamefont {King}, \citenamefont
  {Oktasendra}, \citenamefont {Avis}, \citenamefont {Konicek}, \citenamefont
  {Wadsworth}, \citenamefont {Jusufi}, \citenamefont {Kusumaatmaja},\ and\
  \citenamefont {Yeganeh}}]{natarajan2020predicting}%
  \BibitemOpen
  \bibfield  {author} {\bibinfo {author} {\bibfnamefont {B.}~\bibnamefont
  {Natarajan}}, \bibinfo {author} {\bibfnamefont {A.}~\bibnamefont
  {Jaishankar}}, \bibinfo {author} {\bibfnamefont {M.}~\bibnamefont {King}},
  \bibinfo {author} {\bibfnamefont {F.}~\bibnamefont {Oktasendra}}, \bibinfo
  {author} {\bibfnamefont {S.~J.}\ \bibnamefont {Avis}}, \bibinfo {author}
  {\bibfnamefont {A.~R.}\ \bibnamefont {Konicek}}, \bibinfo {author}
  {\bibfnamefont {G.}~\bibnamefont {Wadsworth}}, \bibinfo {author}
  {\bibfnamefont {A.}~\bibnamefont {Jusufi}}, \bibinfo {author} {\bibfnamefont
  {H.}~\bibnamefont {Kusumaatmaja}},\ and\ \bibinfo {author} {\bibfnamefont
  {M.~S.}\ \bibnamefont {Yeganeh}},\ }\bibfield  {title} {\bibinfo {title}
  {Predicting hemiwicking dynamics on textured substrates},\ }\href@noop {}
  {\bibfield  {journal} {\bibinfo  {journal} {Langmuir}\ }\textbf {\bibinfo
  {volume} {37}},\ \bibinfo {pages} {188} (\bibinfo {year} {2020})}\BibitemShut
  {NoStop}%
\bibitem [{\citenamefont {Courbin}\ \emph {et~al.}(2007)\citenamefont
  {Courbin}, \citenamefont {Denieul}, \citenamefont {Dressaire}, \citenamefont
  {Roper}, \citenamefont {Ajdari},\ and\ \citenamefont
  {Stone}}]{courbin2007imbibition}%
  \BibitemOpen
  \bibfield  {author} {\bibinfo {author} {\bibfnamefont {L.}~\bibnamefont
  {Courbin}}, \bibinfo {author} {\bibfnamefont {E.}~\bibnamefont {Denieul}},
  \bibinfo {author} {\bibfnamefont {E.}~\bibnamefont {Dressaire}}, \bibinfo
  {author} {\bibfnamefont {M.}~\bibnamefont {Roper}}, \bibinfo {author}
  {\bibfnamefont {A.}~\bibnamefont {Ajdari}},\ and\ \bibinfo {author}
  {\bibfnamefont {H.~A.}\ \bibnamefont {Stone}},\ }\bibfield  {title} {\bibinfo
  {title} {Imbibition by polygonal spreading on microdecorated surfaces},\
  }\href@noop {} {\bibfield  {journal} {\bibinfo  {journal} {Nature materials}\
  }\textbf {\bibinfo {volume} {6}},\ \bibinfo {pages} {661} (\bibinfo {year}
  {2007})}\BibitemShut {NoStop}%
\bibitem [{\citenamefont {Chu}\ \emph {et~al.}(2010)\citenamefont {Chu},
  \citenamefont {Xiao},\ and\ \citenamefont {Wang}}]{chu2010uni}%
  \BibitemOpen
  \bibfield  {author} {\bibinfo {author} {\bibfnamefont {K.-H.}\ \bibnamefont
  {Chu}}, \bibinfo {author} {\bibfnamefont {R.}~\bibnamefont {Xiao}},\ and\
  \bibinfo {author} {\bibfnamefont {E.~N.}\ \bibnamefont {Wang}},\ }\bibfield
  {title} {\bibinfo {title} {Uni-directional liquid spreading on asymmetric
  nanostructured surfaces},\ }\href@noop {} {\bibfield  {journal} {\bibinfo
  {journal} {Nature materials}\ }\textbf {\bibinfo {volume} {9}},\ \bibinfo
  {pages} {413} (\bibinfo {year} {2010})}\BibitemShut {NoStop}%
\bibitem [{\citenamefont {Liu}\ \emph {et~al.}(2017)\citenamefont {Liu},
  \citenamefont {Sun}, \citenamefont {Li}, \citenamefont {Xiang}, \citenamefont
  {Che}, \citenamefont {Wang},\ and\ \citenamefont {Zhou}}]{liu2017long}%
  \BibitemOpen
  \bibfield  {author} {\bibinfo {author} {\bibfnamefont {C.}~\bibnamefont
  {Liu}}, \bibinfo {author} {\bibfnamefont {J.}~\bibnamefont {Sun}}, \bibinfo
  {author} {\bibfnamefont {J.}~\bibnamefont {Li}}, \bibinfo {author}
  {\bibfnamefont {C.}~\bibnamefont {Xiang}}, \bibinfo {author} {\bibfnamefont
  {L.}~\bibnamefont {Che}}, \bibinfo {author} {\bibfnamefont {Z.}~\bibnamefont
  {Wang}},\ and\ \bibinfo {author} {\bibfnamefont {X.}~\bibnamefont {Zhou}},\
  }\bibfield  {title} {\bibinfo {title} {Long-range spontaneous droplet
  self-propulsion on wettability gradient surfaces},\ }\href@noop {} {\bibfield
   {journal} {\bibinfo  {journal} {Scientific reports}\ }\textbf {\bibinfo
  {volume} {7}},\ \bibinfo {pages} {7552} (\bibinfo {year} {2017})}\BibitemShut
  {NoStop}%
\bibitem [{\citenamefont {Yang}\ \emph {et~al.}(2024)\citenamefont {Yang},
  \citenamefont {Li}, \citenamefont {Lian}, \citenamefont {Zhu}, \citenamefont
  {Deng}, \citenamefont {Zhang}, \citenamefont {Li}, \citenamefont {Yin},\ and\
  \citenamefont {Wang}}]{yang2024selective}%
  \BibitemOpen
  \bibfield  {author} {\bibinfo {author} {\bibfnamefont {L.}~\bibnamefont
  {Yang}}, \bibinfo {author} {\bibfnamefont {W.}~\bibnamefont {Li}}, \bibinfo
  {author} {\bibfnamefont {J.}~\bibnamefont {Lian}}, \bibinfo {author}
  {\bibfnamefont {H.}~\bibnamefont {Zhu}}, \bibinfo {author} {\bibfnamefont
  {Q.}~\bibnamefont {Deng}}, \bibinfo {author} {\bibfnamefont {Y.}~\bibnamefont
  {Zhang}}, \bibinfo {author} {\bibfnamefont {J.}~\bibnamefont {Li}}, \bibinfo
  {author} {\bibfnamefont {X.}~\bibnamefont {Yin}},\ and\ \bibinfo {author}
  {\bibfnamefont {L.}~\bibnamefont {Wang}},\ }\bibfield  {title} {\bibinfo
  {title} {Selective directional liquid transport on shoot surfaces of crassula
  muscosa},\ }\href@noop {} {\bibfield  {journal} {\bibinfo  {journal}
  {Science}\ }\textbf {\bibinfo {volume} {384}},\ \bibinfo {pages} {1344}
  (\bibinfo {year} {2024})}\BibitemShut {NoStop}%
\bibitem [{\citenamefont {Li}\ \emph {et~al.}(2017)\citenamefont {Li},
  \citenamefont {Zhou}, \citenamefont {Li}, \citenamefont {Che}, \citenamefont
  {Yao}, \citenamefont {McHale}, \citenamefont {Chaudhury},\ and\ \citenamefont
  {Wang}}]{li2017topological}%
  \BibitemOpen
  \bibfield  {author} {\bibinfo {author} {\bibfnamefont {J.}~\bibnamefont
  {Li}}, \bibinfo {author} {\bibfnamefont {X.}~\bibnamefont {Zhou}}, \bibinfo
  {author} {\bibfnamefont {J.}~\bibnamefont {Li}}, \bibinfo {author}
  {\bibfnamefont {L.}~\bibnamefont {Che}}, \bibinfo {author} {\bibfnamefont
  {J.}~\bibnamefont {Yao}}, \bibinfo {author} {\bibfnamefont {G.}~\bibnamefont
  {McHale}}, \bibinfo {author} {\bibfnamefont {M.~K.}\ \bibnamefont
  {Chaudhury}},\ and\ \bibinfo {author} {\bibfnamefont {Z.}~\bibnamefont
  {Wang}},\ }\bibfield  {title} {\bibinfo {title} {Topological liquid diode},\
  }\href@noop {} {\bibfield  {journal} {\bibinfo  {journal} {Science advances}\
  }\textbf {\bibinfo {volume} {3}},\ \bibinfo {pages} {eaao3530} (\bibinfo
  {year} {2017})}\BibitemShut {NoStop}%
\bibitem [{\citenamefont {Cira}\ \emph {et~al.}(2015)\citenamefont {Cira},
  \citenamefont {Benusiglio},\ and\ \citenamefont {Prakash}}]{cira2015vapour}%
  \BibitemOpen
  \bibfield  {author} {\bibinfo {author} {\bibfnamefont {N.~J.}\ \bibnamefont
  {Cira}}, \bibinfo {author} {\bibfnamefont {A.}~\bibnamefont {Benusiglio}},\
  and\ \bibinfo {author} {\bibfnamefont {M.}~\bibnamefont {Prakash}},\
  }\bibfield  {title} {\bibinfo {title} {Vapour-mediated sensing and motility
  in two-component droplets},\ }\href@noop {} {\bibfield  {journal} {\bibinfo
  {journal} {Nature}\ }\textbf {\bibinfo {volume} {519}},\ \bibinfo {pages}
  {446} (\bibinfo {year} {2015})}\BibitemShut {NoStop}%
\bibitem [{\citenamefont {Karpitschka}\ \emph {et~al.}(2017)\citenamefont
  {Karpitschka}, \citenamefont {Liebig},\ and\ \citenamefont
  {Riegler}}]{karpitschka2017marangoni}%
  \BibitemOpen
  \bibfield  {author} {\bibinfo {author} {\bibfnamefont {S.}~\bibnamefont
  {Karpitschka}}, \bibinfo {author} {\bibfnamefont {F.}~\bibnamefont
  {Liebig}},\ and\ \bibinfo {author} {\bibfnamefont {H.}~\bibnamefont
  {Riegler}},\ }\bibfield  {title} {\bibinfo {title} {Marangoni contraction of
  evaporating sessile droplets of binary mixtures},\ }\href@noop {} {\bibfield
  {journal} {\bibinfo  {journal} {Langmuir}\ }\textbf {\bibinfo {volume}
  {33}},\ \bibinfo {pages} {4682} (\bibinfo {year} {2017})}\BibitemShut
  {NoStop}%
\bibitem [{\citenamefont {Baumgartner}\ \emph {et~al.}(2022)\citenamefont
  {Baumgartner}, \citenamefont {Shiri}, \citenamefont {Sinha}, \citenamefont
  {Karpitschka},\ and\ \citenamefont {Cira}}]{baumgartner2022marangoni}%
  \BibitemOpen
  \bibfield  {author} {\bibinfo {author} {\bibfnamefont {D.~A.}\ \bibnamefont
  {Baumgartner}}, \bibinfo {author} {\bibfnamefont {S.}~\bibnamefont {Shiri}},
  \bibinfo {author} {\bibfnamefont {S.}~\bibnamefont {Sinha}}, \bibinfo
  {author} {\bibfnamefont {S.}~\bibnamefont {Karpitschka}},\ and\ \bibinfo
  {author} {\bibfnamefont {N.~J.}\ \bibnamefont {Cira}},\ }\bibfield  {title}
  {\bibinfo {title} {Marangoni spreading and contracting three-component
  droplets on completely wetting surfaces},\ }\href@noop {} {\bibfield
  {journal} {\bibinfo  {journal} {Proceedings of the National Academy of
  Sciences}\ }\textbf {\bibinfo {volume} {119}},\ \bibinfo {pages}
  {e2120432119} (\bibinfo {year} {2022})}\BibitemShut {NoStop}%
\bibitem [{\citenamefont {Yang}\ \emph {et~al.}(2023)\citenamefont {Yang},
  \citenamefont {Pahlavan}, \citenamefont {Stone},\ and\ \citenamefont
  {Bain}}]{yang2023evaporation}%
  \BibitemOpen
  \bibfield  {author} {\bibinfo {author} {\bibfnamefont {L.}~\bibnamefont
  {Yang}}, \bibinfo {author} {\bibfnamefont {A.~A.}\ \bibnamefont {Pahlavan}},
  \bibinfo {author} {\bibfnamefont {H.~A.}\ \bibnamefont {Stone}},\ and\
  \bibinfo {author} {\bibfnamefont {C.~D.}\ \bibnamefont {Bain}},\ }\bibfield
  {title} {\bibinfo {title} {Evaporation of alcohol droplets on surfaces in
  moist air},\ }\href@noop {} {\bibfield  {journal} {\bibinfo  {journal}
  {Proceedings of the National Academy of Sciences}\ }\textbf {\bibinfo
  {volume} {120}},\ \bibinfo {pages} {e2302653120} (\bibinfo {year}
  {2023})}\BibitemShut {NoStop}%
\bibitem [{\citenamefont {Pahlavan}\ \emph {et~al.}(2021)\citenamefont
  {Pahlavan}, \citenamefont {Yang}, \citenamefont {Bain},\ and\ \citenamefont
  {Stone}}]{pahlavan2021evaporation}%
  \BibitemOpen
  \bibfield  {author} {\bibinfo {author} {\bibfnamefont {A.~A.}\ \bibnamefont
  {Pahlavan}}, \bibinfo {author} {\bibfnamefont {L.}~\bibnamefont {Yang}},
  \bibinfo {author} {\bibfnamefont {C.~D.}\ \bibnamefont {Bain}},\ and\
  \bibinfo {author} {\bibfnamefont {H.~A.}\ \bibnamefont {Stone}},\ }\bibfield
  {title} {\bibinfo {title} {Evaporation of binary-mixture liquid droplets: The
  formation of picoliter pancakelike shapes},\ }\href
  {https://doi.org/10.1103/physrevlett.127.024501} {\bibfield  {journal}
  {\bibinfo  {journal} {Physical Review Letters}\ }\textbf {\bibinfo {volume}
  {127}},\ \bibinfo {pages} {024501} (\bibinfo {year} {2021})}\BibitemShut
  {NoStop}%
\bibitem [{\citenamefont {Ishino}\ \emph {et~al.}(2007)\citenamefont {Ishino},
  \citenamefont {Reyssat}, \citenamefont {Reyssat}, \citenamefont {Okumura},\
  and\ \citenamefont {Quere}}]{ishino2007wicking}%
  \BibitemOpen
  \bibfield  {author} {\bibinfo {author} {\bibfnamefont {C.}~\bibnamefont
  {Ishino}}, \bibinfo {author} {\bibfnamefont {M.}~\bibnamefont {Reyssat}},
  \bibinfo {author} {\bibfnamefont {E.}~\bibnamefont {Reyssat}}, \bibinfo
  {author} {\bibfnamefont {K.}~\bibnamefont {Okumura}},\ and\ \bibinfo {author}
  {\bibfnamefont {D.}~\bibnamefont {Quere}},\ }\bibfield  {title} {\bibinfo
  {title} {Wicking within forests of micropillars},\ }\href@noop {} {\bibfield
  {journal} {\bibinfo  {journal} {Europhysics letters}\ }\textbf {\bibinfo
  {volume} {79}},\ \bibinfo {pages} {56005} (\bibinfo {year}
  {2007})}\BibitemShut {NoStop}%
\bibitem [{\citenamefont {Gambaryan-Roisman}(2014)}]{gambaryan2014liquids}%
  \BibitemOpen
  \bibfield  {author} {\bibinfo {author} {\bibfnamefont {T.}~\bibnamefont
  {Gambaryan-Roisman}},\ }\bibfield  {title} {\bibinfo {title} {Liquids on
  porous layers: wetting, imbibition and transport processes},\ }\href@noop {}
  {\bibfield  {journal} {\bibinfo  {journal} {Current Opinion in Colloid \&
  Interface Science}\ }\textbf {\bibinfo {volume} {19}},\ \bibinfo {pages}
  {320} (\bibinfo {year} {2014})}\BibitemShut {NoStop}%
\bibitem [{\citenamefont {Tsoumpas}\ \emph {et~al.}(2015)\citenamefont
  {Tsoumpas}, \citenamefont {Dehaeck}, \citenamefont {Rednikov},\ and\
  \citenamefont {Colinet}}]{tsoumpas2015effect}%
  \BibitemOpen
  \bibfield  {author} {\bibinfo {author} {\bibfnamefont {Y.}~\bibnamefont
  {Tsoumpas}}, \bibinfo {author} {\bibfnamefont {S.}~\bibnamefont {Dehaeck}},
  \bibinfo {author} {\bibfnamefont {A.}~\bibnamefont {Rednikov}},\ and\
  \bibinfo {author} {\bibfnamefont {P.}~\bibnamefont {Colinet}},\ }\bibfield
  {title} {\bibinfo {title} {Effect of marangoni flows on the shape of thin
  sessile droplets evaporating into air},\ }\href@noop {} {\bibfield  {journal}
  {\bibinfo  {journal} {Langmuir}\ }\textbf {\bibinfo {volume} {31}},\ \bibinfo
  {pages} {13334} (\bibinfo {year} {2015})}\BibitemShut {NoStop}%
\bibitem [{\citenamefont {Benusiglio}\ \emph {et~al.}(2018)\citenamefont
  {Benusiglio}, \citenamefont {Cira},\ and\ \citenamefont
  {Prakash}}]{benusiglio2018two}%
  \BibitemOpen
  \bibfield  {author} {\bibinfo {author} {\bibfnamefont {A.}~\bibnamefont
  {Benusiglio}}, \bibinfo {author} {\bibfnamefont {N.~J.}\ \bibnamefont
  {Cira}},\ and\ \bibinfo {author} {\bibfnamefont {M.}~\bibnamefont
  {Prakash}},\ }\bibfield  {title} {\bibinfo {title} {Two-component
  marangoni-contracted droplets: friction and shape},\ }\href@noop {}
  {\bibfield  {journal} {\bibinfo  {journal} {Soft Matter}\ }\textbf {\bibinfo
  {volume} {14}},\ \bibinfo {pages} {7724} (\bibinfo {year}
  {2018})}\BibitemShut {NoStop}%
\bibitem [{\citenamefont {Ram{\'{\i}}rez-Soto}\ and\ \citenamefont
  {Karpitschka}(2022)}]{Ramirez:PRF2022}%
  \BibitemOpen
  \bibfield  {author} {\bibinfo {author} {\bibfnamefont {O.}~\bibnamefont
  {Ram{\'{\i}}rez-Soto}}\ and\ \bibinfo {author} {\bibfnamefont
  {S.}~\bibnamefont {Karpitschka}},\ }\bibfield  {title} {\bibinfo {title}
  {Taylor dispersion in thin liquid films of volatile mixtures: A quantitative
  model for marangoni contraction},\ }\href
  {https://doi.org/10.1103/physrevfluids.7.l022001} {\bibfield  {journal}
  {\bibinfo  {journal} {Physical Review Fluids}\ }\textbf {\bibinfo {volume}
  {7}},\ \bibinfo {pages} {l022001} (\bibinfo {year} {2022})}\BibitemShut
  {NoStop}%
\bibitem [{\citenamefont {Charlier}\ \emph {et~al.}(2022)\citenamefont
  {Charlier}, \citenamefont {Rednikov}, \citenamefont {Dehaeck}, \citenamefont
  {Colinet},\ and\ \citenamefont {Terwagne}}]{charlier2022water}%
  \BibitemOpen
  \bibfield  {author} {\bibinfo {author} {\bibfnamefont {J.}~\bibnamefont
  {Charlier}}, \bibinfo {author} {\bibfnamefont {A.}~\bibnamefont {Rednikov}},
  \bibinfo {author} {\bibfnamefont {S.}~\bibnamefont {Dehaeck}}, \bibinfo
  {author} {\bibfnamefont {P.}~\bibnamefont {Colinet}},\ and\ \bibinfo {author}
  {\bibfnamefont {D.}~\bibnamefont {Terwagne}},\ }\bibfield  {title} {\bibinfo
  {title} {Water--propylene glycol sessile droplet shapes and migration:
  Marangoni mixing and separation of scales},\ }\href@noop {} {\bibfield
  {journal} {\bibinfo  {journal} {Journal of Fluid Mechanics}\ }\textbf
  {\bibinfo {volume} {933}},\ \bibinfo {pages} {A45} (\bibinfo {year}
  {2022})}\BibitemShut {NoStop}%
\bibitem [{\citenamefont {Srivastava}\ \emph {et~al.}(2010)\citenamefont
  {Srivastava}, \citenamefont {Din}, \citenamefont {Judson}, \citenamefont
  {MacDonald},\ and\ \citenamefont {Meinhart}}]{srivastava2010unified}%
  \BibitemOpen
  \bibfield  {author} {\bibinfo {author} {\bibfnamefont {N.}~\bibnamefont
  {Srivastava}}, \bibinfo {author} {\bibfnamefont {C.}~\bibnamefont {Din}},
  \bibinfo {author} {\bibfnamefont {A.}~\bibnamefont {Judson}}, \bibinfo
  {author} {\bibfnamefont {N.~C.}\ \bibnamefont {MacDonald}},\ and\ \bibinfo
  {author} {\bibfnamefont {C.~D.}\ \bibnamefont {Meinhart}},\ }\bibfield
  {title} {\bibinfo {title} {A unified scaling model for flow through a lattice
  of microfabricated posts},\ }\href@noop {} {\bibfield  {journal} {\bibinfo
  {journal} {Lab on a Chip}\ }\textbf {\bibinfo {volume} {10}},\ \bibinfo
  {pages} {1148} (\bibinfo {year} {2010})}\BibitemShut {NoStop}%
\bibitem [{\citenamefont {Kim}\ \emph {et~al.}(2011)\citenamefont {Kim},
  \citenamefont {Moon}, \citenamefont {Lee}, \citenamefont {Mahadevan},\ and\
  \citenamefont {Kim}}]{kim2011hydrodynamics}%
  \BibitemOpen
  \bibfield  {author} {\bibinfo {author} {\bibfnamefont {J.}~\bibnamefont
  {Kim}}, \bibinfo {author} {\bibfnamefont {M.-W.}\ \bibnamefont {Moon}},
  \bibinfo {author} {\bibfnamefont {K.-R.}\ \bibnamefont {Lee}}, \bibinfo
  {author} {\bibfnamefont {L.}~\bibnamefont {Mahadevan}},\ and\ \bibinfo
  {author} {\bibfnamefont {H.-Y.}\ \bibnamefont {Kim}},\ }\bibfield  {title}
  {\bibinfo {title} {Hydrodynamics of writing with ink},\ }\href@noop {}
  {\bibfield  {journal} {\bibinfo  {journal} {Physical review letters}\
  }\textbf {\bibinfo {volume} {107}},\ \bibinfo {pages} {264501} (\bibinfo
  {year} {2011})}\BibitemShut {NoStop}%
\bibitem [{\citenamefont {Krishnan}\ \emph {et~al.}(2019)\citenamefont
  {Krishnan}, \citenamefont {Bal},\ and\ \citenamefont
  {Putnam}}]{krishnan2019simple}%
  \BibitemOpen
  \bibfield  {author} {\bibinfo {author} {\bibfnamefont {S.~R.}\ \bibnamefont
  {Krishnan}}, \bibinfo {author} {\bibfnamefont {J.}~\bibnamefont {Bal}},\ and\
  \bibinfo {author} {\bibfnamefont {S.~A.}\ \bibnamefont {Putnam}},\ }\bibfield
   {title} {\bibinfo {title} {A simple analytic model for predicting the
  wicking velocity in micropillar arrays},\ }\href@noop {} {\bibfield
  {journal} {\bibinfo  {journal} {Scientific reports}\ }\textbf {\bibinfo
  {volume} {9}},\ \bibinfo {pages} {20074} (\bibinfo {year}
  {2019})}\BibitemShut {NoStop}%
\bibitem [{\citenamefont {Courbin}\ \emph {et~al.}(2009)\citenamefont
  {Courbin}, \citenamefont {Bird}, \citenamefont {Reyssat},\ and\ \citenamefont
  {Stone}}]{courbin2009dynamics}%
  \BibitemOpen
  \bibfield  {author} {\bibinfo {author} {\bibfnamefont {L.}~\bibnamefont
  {Courbin}}, \bibinfo {author} {\bibfnamefont {J.~C.}\ \bibnamefont {Bird}},
  \bibinfo {author} {\bibfnamefont {M.}~\bibnamefont {Reyssat}},\ and\ \bibinfo
  {author} {\bibfnamefont {H.~A.}\ \bibnamefont {Stone}},\ }\bibfield  {title}
  {\bibinfo {title} {Dynamics of wetting: from inertial spreading to viscous
  imbibition},\ }\href@noop {} {\bibfield  {journal} {\bibinfo  {journal}
  {Journal of Physics: Condensed Matter}\ }\textbf {\bibinfo {volume} {21}},\
  \bibinfo {pages} {464127} (\bibinfo {year} {2009})}\BibitemShut {NoStop}%
\bibitem [{\citenamefont {Sontheimer}\ \emph {et~al.}(2024)\citenamefont
  {Sontheimer}, \citenamefont {Ho}, \citenamefont {Els{\"a}{\ss}er},
  \citenamefont {Stephan},\ and\ \citenamefont
  {Gambaryan-Roisman}}]{sontheimer2024numerical}%
  \BibitemOpen
  \bibfield  {author} {\bibinfo {author} {\bibfnamefont {H.}~\bibnamefont
  {Sontheimer}}, \bibinfo {author} {\bibfnamefont {A.~T.}\ \bibnamefont {Ho}},
  \bibinfo {author} {\bibfnamefont {L.}~\bibnamefont {Els{\"a}{\ss}er}},
  \bibinfo {author} {\bibfnamefont {P.}~\bibnamefont {Stephan}},\ and\ \bibinfo
  {author} {\bibfnamefont {T.}~\bibnamefont {Gambaryan-Roisman}},\ }\bibfield
  {title} {\bibinfo {title} {Numerical simulation of drop impingement onto
  superheated textured walls},\ }in\ \href@noop {} {\emph {\bibinfo {booktitle}
  {Journal of Physics: Conference Series}}},\ Vol.\ \bibinfo {volume} {2766}\
  (\bibinfo {organization} {IOP Publishing},\ \bibinfo {year} {2024})\ p.\
  \bibinfo {pages} {012085}\BibitemShut {NoStop}%
\bibitem [{\citenamefont {Lou}\ \emph {et~al.}(2022)\citenamefont {Lou},
  \citenamefont {Shi}, \citenamefont {Ma}, \citenamefont {Zhou}, \citenamefont
  {Huang}, \citenamefont {Zheng},\ and\ \citenamefont {Lv}}]{lou2022polygonal}%
  \BibitemOpen
  \bibfield  {author} {\bibinfo {author} {\bibfnamefont {J.}~\bibnamefont
  {Lou}}, \bibinfo {author} {\bibfnamefont {S.}~\bibnamefont {Shi}}, \bibinfo
  {author} {\bibfnamefont {C.}~\bibnamefont {Ma}}, \bibinfo {author}
  {\bibfnamefont {X.}~\bibnamefont {Zhou}}, \bibinfo {author} {\bibfnamefont
  {D.}~\bibnamefont {Huang}}, \bibinfo {author} {\bibfnamefont
  {Q.}~\bibnamefont {Zheng}},\ and\ \bibinfo {author} {\bibfnamefont
  {C.}~\bibnamefont {Lv}},\ }\bibfield  {title} {\bibinfo {title} {Polygonal
  non-wetting droplets on microtextured surfaces},\ }\href@noop {} {\bibfield
  {journal} {\bibinfo  {journal} {Nature Communications}\ }\textbf {\bibinfo
  {volume} {13}},\ \bibinfo {pages} {2685} (\bibinfo {year}
  {2022})}\BibitemShut {NoStop}%
\bibitem [{\citenamefont {Raj}\ \emph {et~al.}(2014)\citenamefont {Raj},
  \citenamefont {Adera}, \citenamefont {Enright},\ and\ \citenamefont
  {Wang}}]{raj2014high}%
  \BibitemOpen
  \bibfield  {author} {\bibinfo {author} {\bibfnamefont {R.}~\bibnamefont
  {Raj}}, \bibinfo {author} {\bibfnamefont {S.}~\bibnamefont {Adera}}, \bibinfo
  {author} {\bibfnamefont {R.}~\bibnamefont {Enright}},\ and\ \bibinfo {author}
  {\bibfnamefont {E.~N.}\ \bibnamefont {Wang}},\ }\bibfield  {title} {\bibinfo
  {title} {High-resolution liquid patterns via three-dimensional droplet shape
  control},\ }\href@noop {} {\bibfield  {journal} {\bibinfo  {journal} {Nature
  communications}\ }\textbf {\bibinfo {volume} {5}},\ \bibinfo {pages} {4975}
  (\bibinfo {year} {2014})}\BibitemShut {NoStop}%
\bibitem [{\citenamefont {Al~Balushi}\ \emph {et~al.}(2023)\citenamefont
  {Al~Balushi}, \citenamefont {Duursma}, \citenamefont {Valluri}, \citenamefont
  {Sefiane},\ and\ \citenamefont {Orejon}}]{al2023binary}%
  \BibitemOpen
  \bibfield  {author} {\bibinfo {author} {\bibfnamefont {K.~M.}\ \bibnamefont
  {Al~Balushi}}, \bibinfo {author} {\bibfnamefont {G.}~\bibnamefont {Duursma}},
  \bibinfo {author} {\bibfnamefont {P.}~\bibnamefont {Valluri}}, \bibinfo
  {author} {\bibfnamefont {K.}~\bibnamefont {Sefiane}},\ and\ \bibinfo {author}
  {\bibfnamefont {D.}~\bibnamefont {Orejon}},\ }\bibfield  {title} {\bibinfo
  {title} {Binary mixture droplet evaporation on microstructured decorated
  surfaces and the mixed stick--slip modes},\ }\href@noop {} {\bibfield
  {journal} {\bibinfo  {journal} {Langmuir}\ }\textbf {\bibinfo {volume}
  {39}},\ \bibinfo {pages} {8323} (\bibinfo {year} {2023})}\BibitemShut
  {NoStop}%
\bibitem [{\citenamefont {Wong}\ \emph {et~al.}(2011)\citenamefont {Wong},
  \citenamefont {Kang}, \citenamefont {Tang}, \citenamefont {Smythe},
  \citenamefont {Hatton}, \citenamefont {Grinthal},\ and\ \citenamefont
  {Aizenberg}}]{Wong:N2011}%
  \BibitemOpen
  \bibfield  {author} {\bibinfo {author} {\bibfnamefont {T.-S.}\ \bibnamefont
  {Wong}}, \bibinfo {author} {\bibfnamefont {S.~H.}\ \bibnamefont {Kang}},
  \bibinfo {author} {\bibfnamefont {S.~K.~Y.}\ \bibnamefont {Tang}}, \bibinfo
  {author} {\bibfnamefont {E.~J.}\ \bibnamefont {Smythe}}, \bibinfo {author}
  {\bibfnamefont {B.~D.}\ \bibnamefont {Hatton}}, \bibinfo {author}
  {\bibfnamefont {A.}~\bibnamefont {Grinthal}},\ and\ \bibinfo {author}
  {\bibfnamefont {J.}~\bibnamefont {Aizenberg}},\ }\bibfield  {title} {\bibinfo
  {title} {Bioinspired self-repairing slippery surfaces with pressure-stable
  omniphobicity},\ }\href {https://doi.org/10.1038/nature10447} {\bibfield
  {journal} {\bibinfo  {journal} {Nature}\ }\textbf {\bibinfo {volume} {477}},\
  \bibinfo {pages} {443} (\bibinfo {year} {2011})}\BibitemShut {NoStop}%
\bibitem [{\citenamefont {Tan}\ \emph {et~al.}(2023)\citenamefont {Tan},
  \citenamefont {Lohse},\ and\ \citenamefont {Zhang}}]{tan2023self}%
  \BibitemOpen
  \bibfield  {author} {\bibinfo {author} {\bibfnamefont {H.}~\bibnamefont
  {Tan}}, \bibinfo {author} {\bibfnamefont {D.}~\bibnamefont {Lohse}},\ and\
  \bibinfo {author} {\bibfnamefont {X.}~\bibnamefont {Zhang}},\ }\bibfield
  {title} {\bibinfo {title} {Self-lubricating drops},\ }\href@noop {}
  {\bibfield  {journal} {\bibinfo  {journal} {Current Opinion in Colloid \&
  Interface Science}\ }\textbf {\bibinfo {volume} {68}},\ \bibinfo {pages}
  {101744} (\bibinfo {year} {2023})}\BibitemShut {NoStop}%
\bibitem [{\citenamefont {Wenzel}(1936)}]{wenzel1936resistance}%
  \BibitemOpen
  \bibfield  {author} {\bibinfo {author} {\bibfnamefont {R.~N.}\ \bibnamefont
  {Wenzel}},\ }\bibfield  {title} {\bibinfo {title} {Resistance of solid
  surfaces to wetting by water},\ }\href@noop {} {\bibfield  {journal}
  {\bibinfo  {journal} {Industrial \& engineering chemistry}\ }\textbf
  {\bibinfo {volume} {28}},\ \bibinfo {pages} {988} (\bibinfo {year}
  {1936})}\BibitemShut {NoStop}%
\bibitem [{\citenamefont {Onda}\ \emph {et~al.}(1996)\citenamefont {Onda},
  \citenamefont {Shibuichi}, \citenamefont {Satoh},\ and\ \citenamefont
  {Tsujii}}]{onda1996super}%
  \BibitemOpen
  \bibfield  {author} {\bibinfo {author} {\bibfnamefont {T.}~\bibnamefont
  {Onda}}, \bibinfo {author} {\bibfnamefont {S.}~\bibnamefont {Shibuichi}},
  \bibinfo {author} {\bibfnamefont {N.}~\bibnamefont {Satoh}},\ and\ \bibinfo
  {author} {\bibfnamefont {K.}~\bibnamefont {Tsujii}},\ }\bibfield  {title}
  {\bibinfo {title} {Super-water-repellent fractal surfaces},\ }\href@noop {}
  {\bibfield  {journal} {\bibinfo  {journal} {Langmuir}\ }\textbf {\bibinfo
  {volume} {12}},\ \bibinfo {pages} {2125} (\bibinfo {year}
  {1996})}\BibitemShut {NoStop}%
\bibitem [{\citenamefont {Shibuichi}\ \emph {et~al.}(1996)\citenamefont
  {Shibuichi}, \citenamefont {Onda}, \citenamefont {Satoh},\ and\ \citenamefont
  {Tsujii}}]{shibuichi1996super}%
  \BibitemOpen
  \bibfield  {author} {\bibinfo {author} {\bibfnamefont {S.}~\bibnamefont
  {Shibuichi}}, \bibinfo {author} {\bibfnamefont {T.}~\bibnamefont {Onda}},
  \bibinfo {author} {\bibfnamefont {N.}~\bibnamefont {Satoh}},\ and\ \bibinfo
  {author} {\bibfnamefont {K.}~\bibnamefont {Tsujii}},\ }\bibfield  {title}
  {\bibinfo {title} {Super water-repellent surfaces resulting from fractal
  structure},\ }\href@noop {} {\bibfield  {journal} {\bibinfo  {journal} {The
  Journal of Physical Chemistry}\ }\textbf {\bibinfo {volume} {100}},\ \bibinfo
  {pages} {19512} (\bibinfo {year} {1996})}\BibitemShut {NoStop}%
\bibitem [{\citenamefont {Mujiburohman}\ \emph {et~al.}(2006)\citenamefont
  {Mujiburohman}, \citenamefont {Sediawan},\ and\ \citenamefont
  {Sulistyo}}]{mujiburohman2006preliminary}%
  \BibitemOpen
  \bibfield  {author} {\bibinfo {author} {\bibfnamefont {M.}~\bibnamefont
  {Mujiburohman}}, \bibinfo {author} {\bibfnamefont {W.~B.}\ \bibnamefont
  {Sediawan}},\ and\ \bibinfo {author} {\bibfnamefont {H.}~\bibnamefont
  {Sulistyo}},\ }\bibfield  {title} {\bibinfo {title} {A preliminary study:
  Distillation of isopropanol--water mixture using fixed adsorptive
  distillation method},\ }\href@noop {} {\bibfield  {journal} {\bibinfo
  {journal} {Separation and purification technology}\ }\textbf {\bibinfo
  {volume} {48}},\ \bibinfo {pages} {85} (\bibinfo {year} {2006})}\BibitemShut
  {NoStop}%
\bibitem [{\citenamefont {Brunjes}\ and\ \citenamefont
  {Bogart}(1943)}]{brunjes1943vapor}%
  \BibitemOpen
  \bibfield  {author} {\bibinfo {author} {\bibfnamefont {A.~S.}\ \bibnamefont
  {Brunjes}}\ and\ \bibinfo {author} {\bibfnamefont {M.~J.}\ \bibnamefont
  {Bogart}},\ }\bibfield  {title} {\bibinfo {title} {Vapor-liquid equilibria
  for commercially important systems of organic solvents: The binary systems
  ethanol-n-butanol, acetone-water and isopropanol-water},\ }\href@noop {}
  {\bibfield  {journal} {\bibinfo  {journal} {Industrial \& Engineering
  Chemistry}\ }\textbf {\bibinfo {volume} {35}},\ \bibinfo {pages} {255}
  (\bibinfo {year} {1943})}\BibitemShut {NoStop}%
\end{thebibliography}

%

\clearpage
\onecolumngrid

\appendix 

\section*{Supplementary information}

\subsection{Supplementary Videos}

\noindent\textbf{Supplementary Video 1: Droplet-film coexistence on pillar decorated surface in IPA atmosphere.}\\
Video for Fig.~\ref{fig:experiment}\textbf{b} of the main manuscript. Depositing a drop of pure water onto a pillar-decorated surface (pillar spacing $S = \SI{30}{\micro\meter}$, pillar dimensions $W = H = \SI{10}{\micro\meter}$, square lattice) in an IPA vapor environment ($\sim 80\%$ saturation) causes a vapor mediated wicking delay and a prolonged coexistence of drop and wicking film.\\

\noindent\textbf{Supplementary Video 2: Impact of pillar spacing.}\\
Video for Fig.~\ref{fig:phasediagram}\textbf{a} of the main manuscript. The pillar spacing governs the wicking force, leading to different coexistence durations for water droplets on pillar-decorated surfaces in IPA atmosphere ($S = \SI{30}{\micro\meter}$ and $\SI{50}{\micro\meter}$, $W = H = \SI{10}{\micro\meter}$).\\

\noindent\textbf{Supplementary Video 3: Vapor-triggered reversible imbibition and extraction.}\\
Video for Fig.~\ref{fig:manipulation}\textbf{a} of the main manuscript. Depositing a water drop in water vapor imbibes into the pillar texture ($S = \SI{30}{\micro\meter}$, $W = H = \SI{10}{\micro\meter}$). Switching the atmosphere to IPA vapor extracts the liquid from the texture into a contracted droplet. Switching the atmosphere back to water vapor imbibes the liquid again, and the process can be repeated.\\

\noindent\textbf{Supplementary Video 4: Impact of lattice symmetry and pillar shape.}\\
Video for Fig.~\ref{fig:manipulation}\textbf{b} of the main manuscript. Drop-film coexistence and film shape are modulated by texture pattern and pillar shape. Anisotropic lattices generate different conditions along axes with small and large pillar spacing, respectively. Coexistence pertains to random patterns and pillar shapes.\\

\noindent\textbf{Supplementary Video 5: Drop motion induced by a pillar spacing gradient.}\\
Video for Fig.~\ref{fig:manipulation}\textbf{c} of the main manuscript. Pillars are uniformly spaced on the $x$-axis ($S_x = \SI{30}{\micro\meter}$) but exhibit a gradient in spacing on the $y$-axis ($S_y = \SIrange{35}{15}{\micro\meter}$, bottom to top). The gradient in wicking force leads to a preferential wicking direction towards smaller $S_y$, dragging the droplet along.\\

\noindent\textbf{Supplementary Video 6: Droplet manipulation on pillar decorated surface by vapor gradient.}\\
Video for Fig.~\ref{fig:manipulation}\textbf{d},~\textbf{e} of the main manuscript. Drop-film coexistence on a pillar-decorated surface (pillar spacing $S = \SI{30}{\micro\meter}$, pillar dimensions $W = H = \SI{10}{\micro\meter}$, square lattice). Droplet motion is induced by a vapor composition gradient (IPA-rich on the left, water-rich on the right). Bounding the pattern to an {\sf S}-shaped curve guides the droplet on a complex path.\\

\noindent\textbf{Supplementary Video 7: Water droplet in water atmosphere on pillar decorated surface.}\\
In vapor of the same kind, immediate imbibition of the liquid is observed ($S = \SI{30}{\micro\meter}$, $W = H = \SI{10}{\micro\meter}$).\\

\noindent\textbf{Supplementary Video 8: Depinning process.}\\
On a patterned surface with large pillar spacings ($S = \SI{50}{\micro\meter}$, $W = H = \SI{10}{\micro\meter}$), imbibition is initially fully inhibited. As contraction weakens over time, a film emerges from the droplet in a zipping-like motion. The first part of the video has been accelerated, part of the process is replayed decelerated in the second part.\\

\clearpage

\subsection{Extended Data Figures}

\setcounter{figure}{0}
\renewcommand{\figurename}{\textbf{Extended Data Fig.}}
\renewcommand{\thefigure}{\textbf{S\arabic{figure}}}

\begin{figure*}[h]
\centering
\includegraphics[scale=1]{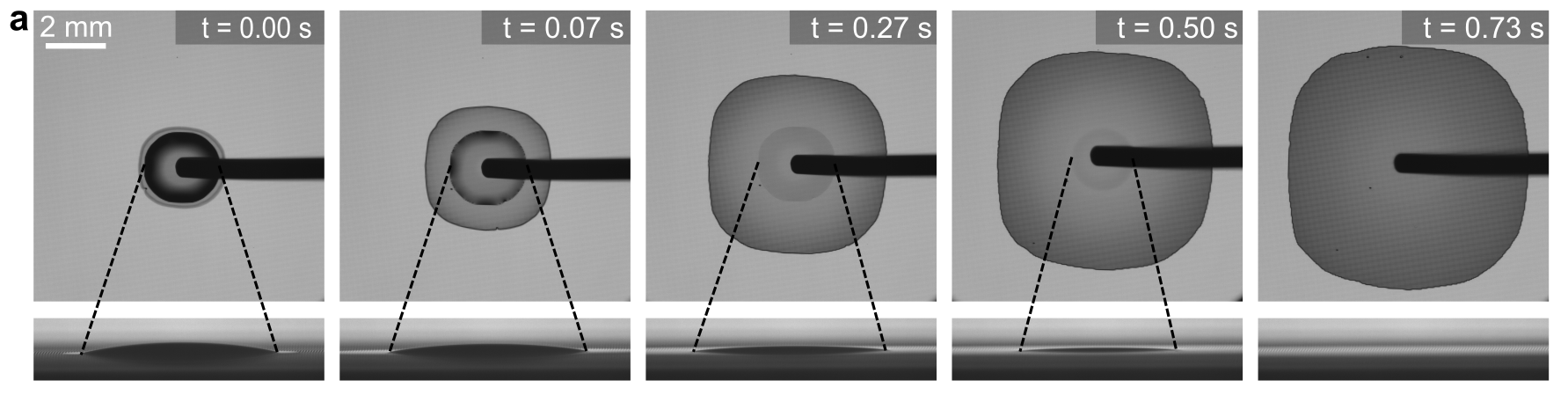}
\caption{Immediate wicking of a $\SI{0.5}{\micro\litre}$ water droplet on a micropillar decorated substrate in an atmosphere with high relative humidity (RH = 84 \%; square lattice with $S=\SI{30}{\micro\meter}$).}
\label{SIfig:immediate}
\end{figure*}

\begin{figure*}[h]
\centering
\includegraphics[scale=1]{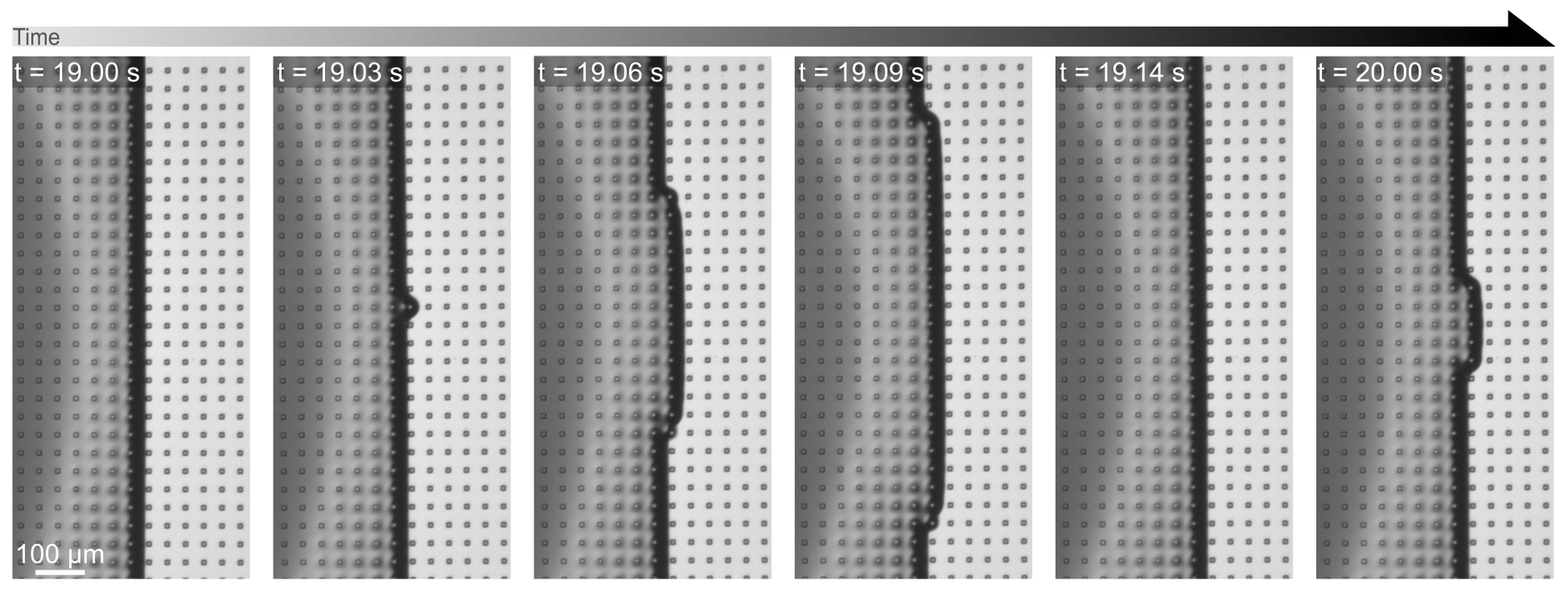}
\caption{Zipping behavior of the imbibition film on a patterned surface (IPA vapor atmosphere; square lattice with $S=\SI{30}{\micro\meter}$).}
\label{SIfig:zipping}
\end{figure*}

\begin{figure*}[h]
\centering
\includegraphics[scale=1]{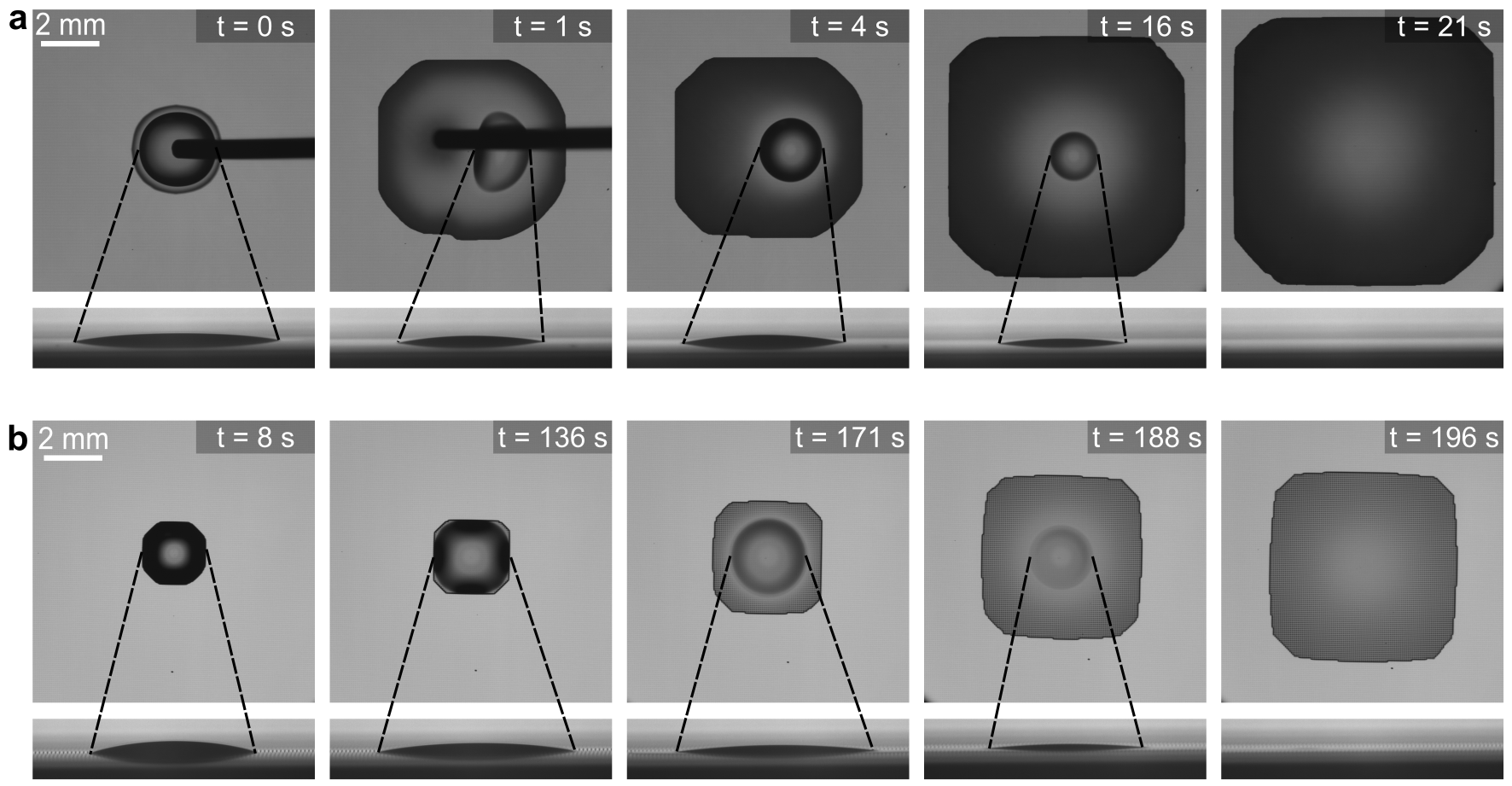}
\caption{Delayed wicking (\textbf{a}, $S= \SI{10}{\micro\meter}$) and initial pinning (\textbf{b}, $S=\SI{50}{\micro\meter}$) of an $\SI{0.5}{\micro\litre}$ water droplet on a square micropillar lattice, in an atmosphere close to saturation with IPA vapor. Mind the different time scales (times indicated top-left).}
\label{SIfig:delayedP10P50}
\end{figure*}

\begin{figure*}[h]
\centering
\includegraphics[scale=1]{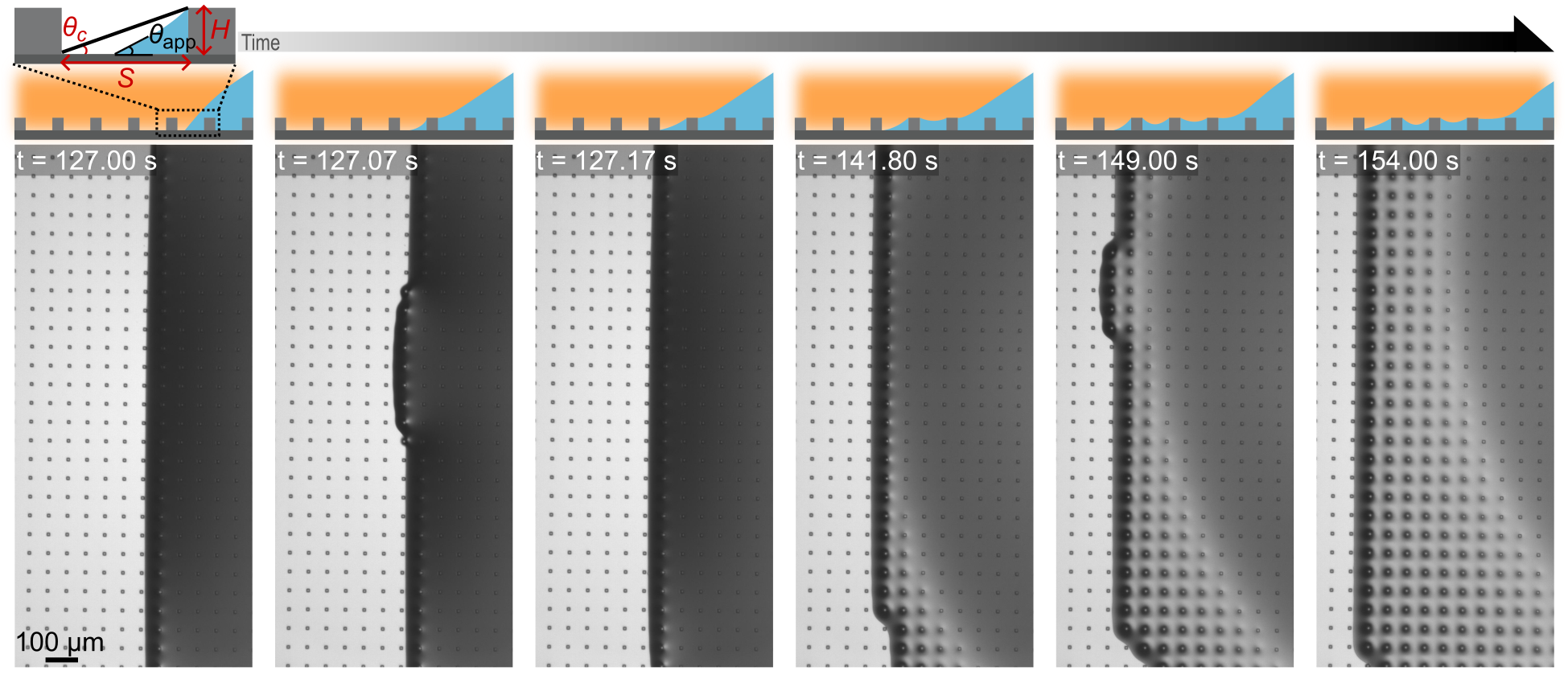}
\caption{Depinning process of an initially pinned drop on a square pillar lattice with large spacing ($S = \SI{50}{\micro\meter}$, $W = H = \SI{10}{\micro\meter}$).}
\label{SIfig:depinning}
\end{figure*}

\begin{figure*}
    \centering
    \includegraphics{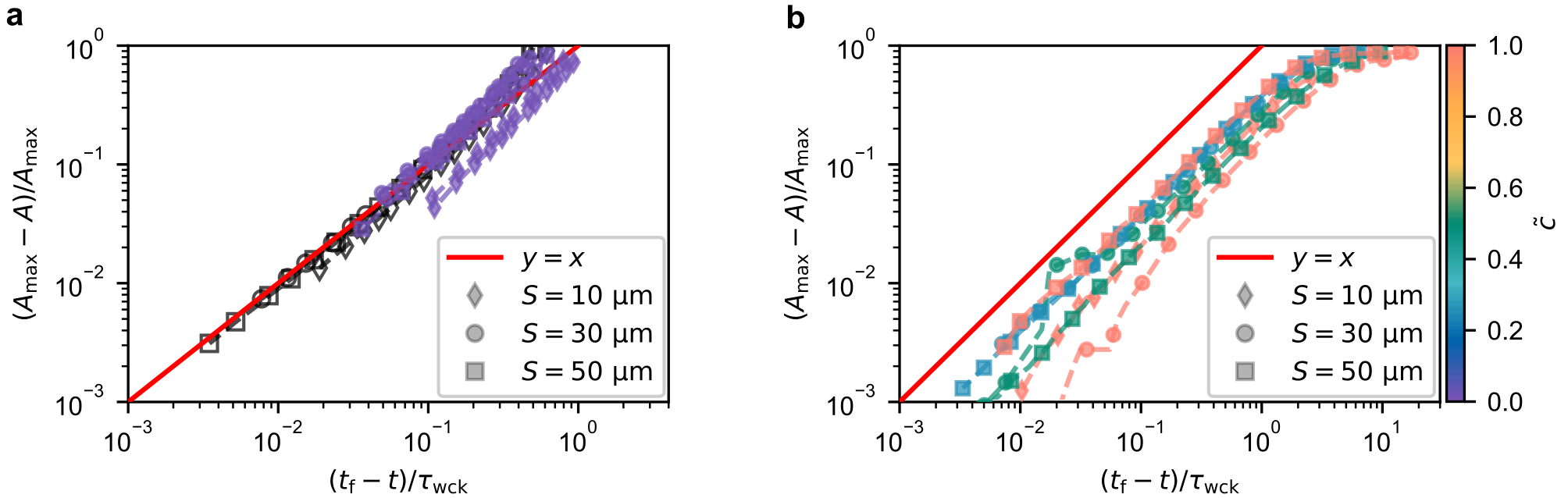}
    \caption{Rescaled film area $( A_{\mathrm{max}} - A)/ A_{\mathrm{max}} $ as function of rescaled time $ (t_{\mathrm{f}} – t)/\tau_{wck} $, with $t_{\mathrm{f}} $ the time at which the contracted droplet totally merged into the wetting film, $A_{\mathrm{max}}$ the film area at that moment, $\tau_{wck} = 2 A_{\mathrm{max}} / \pi D_{w}$ the theoretical wicking time scale (see SI section \emph{Wicking dynamics}).
    The markers correspond to different pillar spacings and the colors correspond to the different liquid IPA mass fractions $\tilde{c}$ inside the bubbler. 
    Panel~\textbf{a} shows `pure' cases, i.e. water droplets in water vapor and IPA droplets IPA vapor, indicated in purple and by black points, respectively, together with the theoretical expectation (red line).
    All data collapse, apart from a small deviation for pure water on narrow pillar spacings, most likely caused by pinning effects.
    Panel~\textbf{b} shows the Marangoni contracted cases, which follow the same power law as the theoretical expectation for pure cases (red line), but with a significantly smaller prefactor.
    }
    \label{SIfig:Film_Wicking_experiments}
\end{figure*}

\begin{figure*}[h]
\centering
\includegraphics[scale=1]{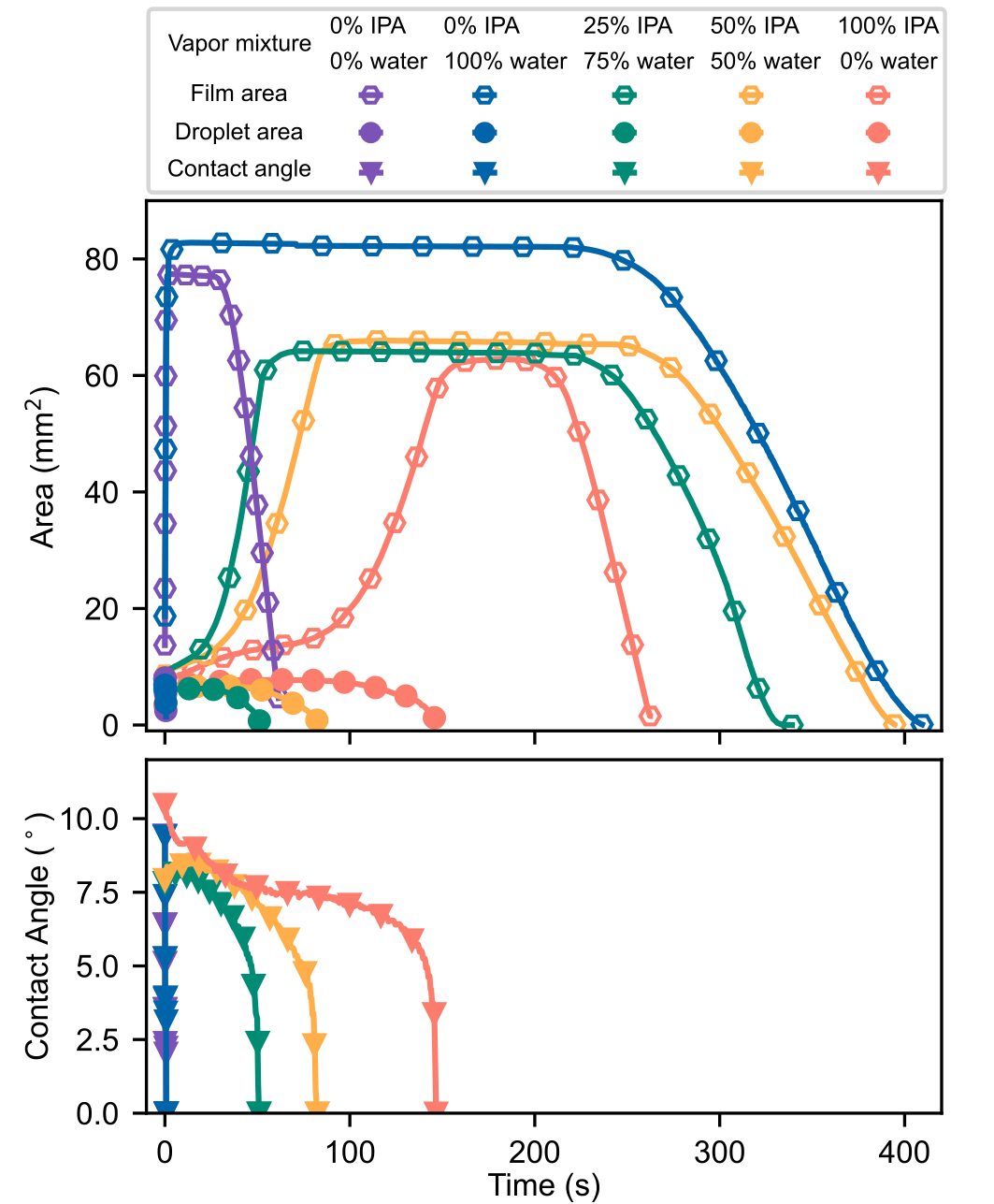}
\caption{Evolution of droplet footprint and wicking film area (top) and apparent contact angle (bottom) on textured surfaces ($S = \SI{30}{\micro\meter}$, $W = H = \SI{10}{\micro\meter}$, square lattice) in different vapor compositions, denoted by the liquid IPA mass fraction in the bubbler.}
\label{SIfig:evolP30}
\end{figure*}

\begin{figure*}[h]
    \centering
    \includegraphics{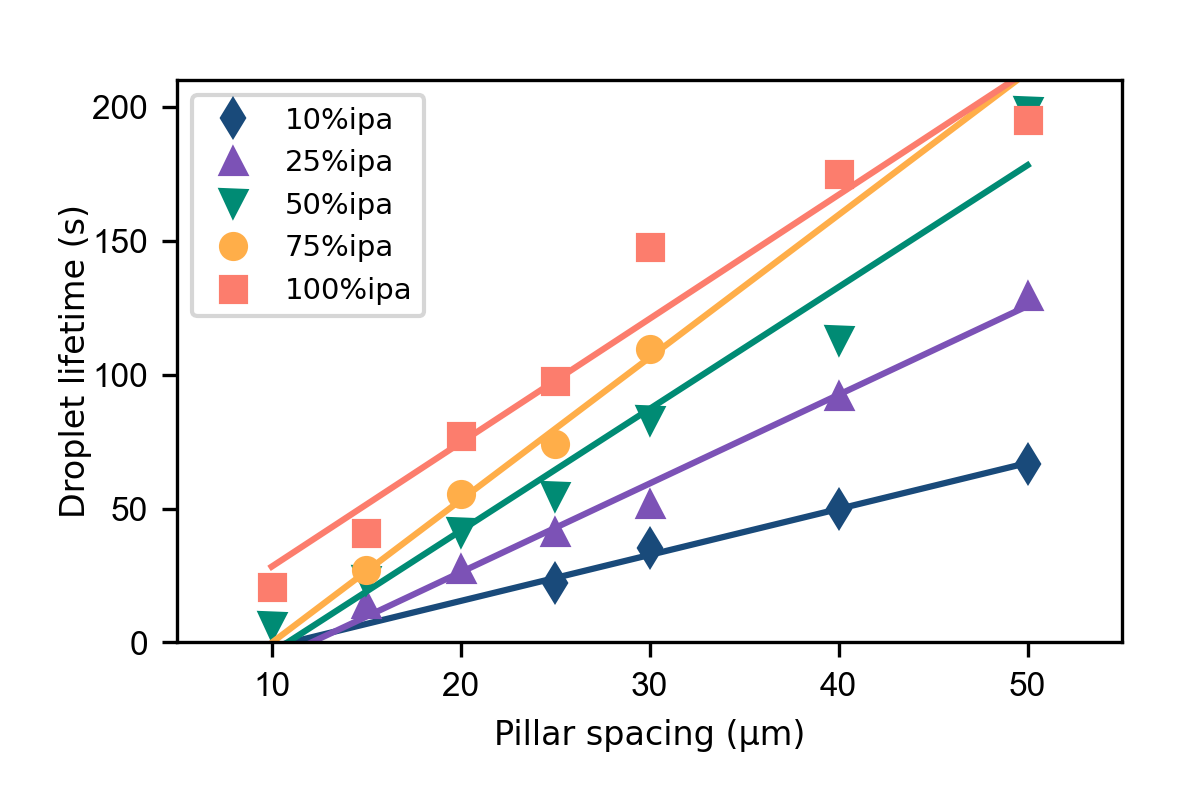}
    \caption{Droplet lifetime as function of the pillar spacing for different IPA mass fraction in the liquid bubbler. The lines are linear fits to each group of data, intended as a guide to the eye.}
    \label{SIfig:Drop_lifetime}
\end{figure*}

\begin{figure*}[h]
\centering
\includegraphics[scale=1]{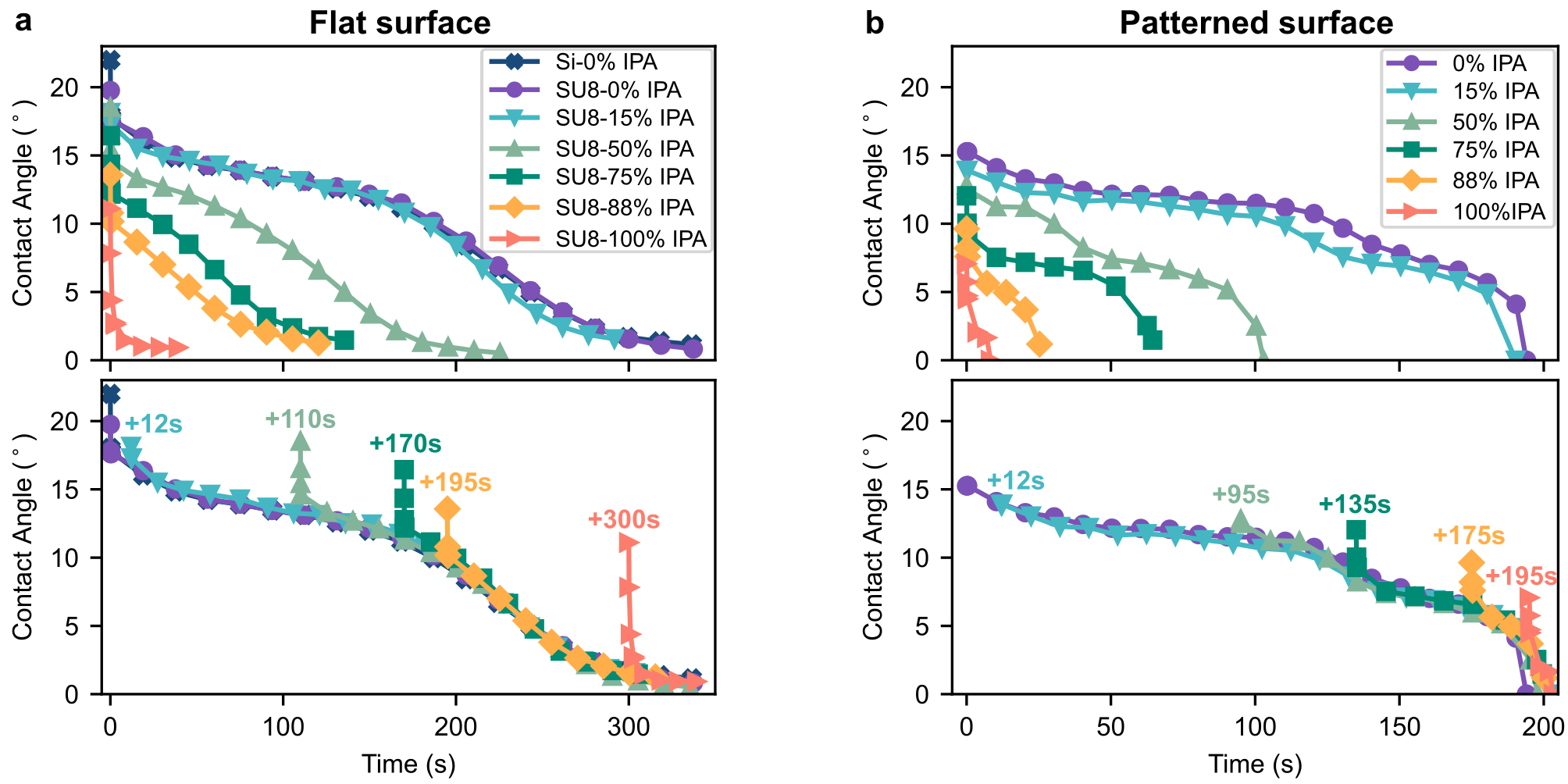}
\caption{Evolution of the apparent contact angle  for droplets with different initial IPA mass fractions in pure IPA vapor environment. \textbf{a}, on flat silicon and flat SU-8 surfaces and \textbf{b}, on patterned surfaces ($S = \SI{50}{\micro\meter}$, $W = H = \SI{10}{\micro\meter}$, square lattice).
The top panels show the data as a function of time since droplet deposition. On the lower panels, data have been collapsed by time-shifting, suggesting a quasi-stationary mechanism that is rather insensitive to the droplet size.
Comparing panels \textbf{a} and \textbf{b} reveals that the droplet lifetime is reduced by the wicking force of the pillars, and that the contraction phase ends more abruptly for the case with pillars.
}
\label{SIfig:flat&p50}
\end{figure*}

\FloatBarrier

\subsection{Wicking dynamics}
\label{sec:Wicking_dynamics}
When a surface is textured with topographic features below the capillary length, surface related effects are enhanced by the roughness factor $r$, defined as the ratio between actual and projected surface area.
This leads to unusual capillary phenomena~\cite{wenzel1936resistance,onda1996super,shibuichi1996super}: for example, hydrophilic or hydrophobic surfaces may be turned into superhydrophilic or superhydrophobic surfaces, respectively.
Whether a partial wetting droplet is imbibed into a pillar array can be determined from an energetic argument, depending only on the Young contact angle $\theta$ at the liquid-vapor-solid contact line, the roughness factor $r = 1 + 4WH / (W+S)^2$, and the pillar density (area fraction) $\phi_{p} = W^2/(S+W)^2$~\cite{bico2002wetting,quere2008wetting}. Namely, if Young's angle $\theta<\theta_c$, where
\begin{equation}
\cos(\theta_{c}) = \frac{1-\phi_{p}}{r-\phi_{p}},
 \end{equation}
wicking occurs. The capillary wicking force per unit length of the contact line, $ F_{w,c} = \gamma (r-\phi_{p}) \left[\cos(\theta) - \cos(\theta_{c}) \right]$ is positive for $\cos(\theta) > \cos(\theta_{c})$, where $\gamma$ is the liquid-vapor surface tension.
Note that geometry enters $F_{w,c}$ and $\theta_c$ solely through ratios of $H$, $W$, and $S$, rendering it fundamentally scale-independent.

The imbibition dynamics of the pillar array is usually called hemi-wicking.
For small scale textures, it is driven by a balance between $F_{w,c}$ and a viscous force per unit length contact line, $F_{\eta}\sim\eta V L$ with $\eta$ the liquid viscosity, $V$ the flow velocity and $L$ the distance between the liquid reservoir and the imbibition front.
This causes the wicking film size to grow according to a diffusion-like behavior $L^{2} \sim D_{w} t$ where $D_{w}$ is a wicking rate constant.
To obtain the exact expression of the wicking constant, one needs to solve the full Stokes equations for flows through the pillar array.
This generates a non-universal behavior as the three dimensions of the problem, the pillar height $H$, width $W$ and spacing $S$ become comparable.
In the work of Natarajan and coworkers~\cite{natarajan2020predicting}, 8 different expressions of $D_{w}$ are listed~\cite{bico2002wetting,ishino2007wicking,srivastava2010unified,kim2011hydrodynamics,kim2016dynamics,krishnan2019simple}.
Typically, viscous resistance depends dominantly on the smallest dimension of the imbibed pattern geometry i.e., either the pillar height $H$ or the pillar spacing $S$.
For $W \ll S$, two limiting cases are found: for tall pillars, $D_{w} \sim \gamma W /\eta$ up to an Onseen logarithmic factor, while for short pillars $D_{w} \sim (\gamma W / \eta) \left(H / (W+S)\right)^{2}$~\cite{quere2008wetting}.
Our experiments with $S = \SI{50}{\micro\meter}$ should typically fall into the latter category, while $S = \SI{30}{\micro\meter}$ might already enter an intermediate regime.

We compare the final wicking phase for ceasing Marangoni contraction toward the end of the droplet lifetime with the expected wicking regimes in Fig.~\ref{SIfig:Film_Wicking_experiments}.
The growth in the film area $A(t) = A_{\mathrm{max}} + \pi D_{w} \left(t_{\mathrm{f}} - t \right)$ with $A_{\mathrm{max}} $ and $t_{\mathrm{f}}$ the wetted area and the time when the contracted droplet completely merged into the film.
We assume that the wicking film is in equilibrium with the gas phase, meaning its surface tension and viscosity correspond to those of a liquid mixture with the same IPA fraction as in the liquid bubbler (see section \emph{IPA-water liquid-vapor equilibrium properties} below for more details).
Experiments of the pure cases -- whether IPA or water droplets in vapor of the same kind -- exhibit wicking dynamics that collapses when using the wicking coefficient:
\begin{equation}
    D_{w} = \frac{2\gamma H}{3\eta} \left( \frac{ 1 -  \cos(\theta_{c})}{\cos(\theta_{c})} \right) \left[ 1-\frac{4W}{\pi U} \tanh{\left(\frac{\pi U}{4W} \right)}\right]
\end{equation}
 with $U = S+SW/(S+W)$ proposed by Natarajan and coworkers~\cite{natarajan2020predicting}.
Notably, only the case of water in water vapor with a pillar spacing of $S=\SI{10}{\micro\meter}$ displays distinctly different behavior, which we attribute to water contact line pinning on SU-8 pillars.
Surprisingly, all our experiments reveal a clear diffusion-like spreading behavior, though the wicking coefficient is significantly lower for cases with Marangoni contraction.

Interestingly, this provides an additional level of control over the descibed phenomenon, as the coexistence regime is governed solely by the balance between the Marangoni force and $F_{w,c}$, which depends only on the geometric ratios.
In contrast, the timescale of wicking and thus the final spreading regime is influenced by the smallest confining size, here $H$, through viscous dissipation.

\subsection{IPA-water liquid-vapor equilibrium properties}
\label{sec:IPA-Water_equilibrium}

\begin{figure*}
    \centering
    \includegraphics{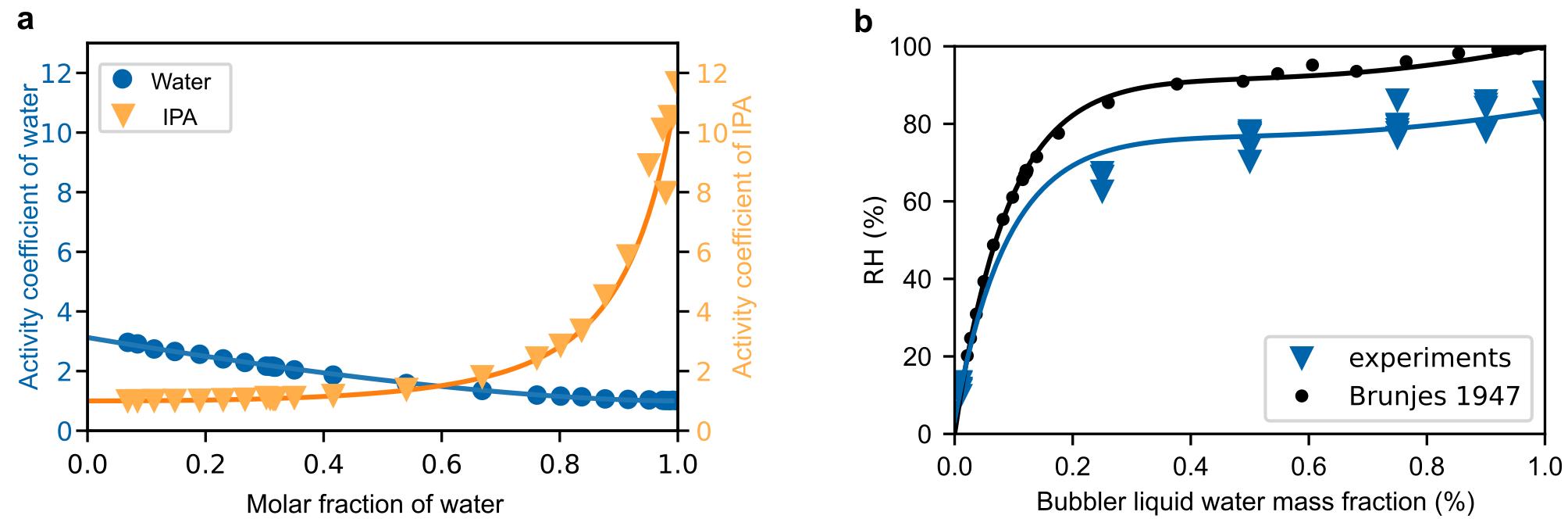}
    \caption{\textbf{a}, Activity of IPA and water of an IPA-water mixture in function of the molar fraction of water in the mixture. The activity values are fitted by Van Laar equations \eqref{Eq:Van_Laar_equations} with fitted parameters B = 2.45 and C = 2.15. \textbf{b}, Measured relative humidity in the chamber compared to a gas phase in equilibrium with an IPA-water liquid mixture.  Data are taken from~\cite{brunjes1943vapor}.}
    \label{SIfig:Activity_RH_IPA-water}
\end{figure*}

IPA and water form a positive azeotrope with a boiling point minimum at $80.4 ^{\circ} C$ and an IPA mass fraction $\tilde{c}\sim 87.7 \%$~\cite{mujiburohman2006preliminary}.
Figure~\ref{SIfig:Activity_RH_IPA-water}\textbf{a} shows the activity coefficients of water and IPA as a function of the molar fraction of water.
The literature values from~\cite{brunjes1943vapor} are well fitted by the so-called Van Laar equations
\begin{equation}
A_{\mathrm{IPA}} = \exp \left( \frac{B}{\left[1+ C x/(1- x) \right]^2} \right), \quad\quad\quad A_{\mathrm{water}} = \exp\left( \frac{BC}{\left[C+ (1-x)/ x \right]^2} \right),
\label{Eq:Van_Laar_equations}
\end{equation}
with fitted parameters $B = 2.45$ and $C = 2.15$.

Under the assumption of diffusion limited evaporation/condensation in an infinitely extended atmosphere with prescribed ambient vapor pressures, the evolution of the total mass $m_{\textbf{D},i}$ of component $i$ in a spherical-cap-shaped droplet scales as
\begin{equation}
\frac{d m_{\mathrm{D},i}}{dt} \sim R M_i D_{g,i} p_{sat,i} \left(A_{i} x_{\mathrm{D},i}  - \frac{p_{v,i}}{ p_{sat,i}}\right),
\label{Eq:IPA_Conservation_equation}
\end{equation}
where $R$ is the droplet footprint radius, $M_i$ is the molar mass, $ D_{g,i} $ is the gas diffusion constant, $x_{\mathrm{D},i}$ is the molar fraction of component $i$ in the droplet, $p_{v,i}$ is the partial pressure, and $p_{sat,i}$ is the saturation vapor pressure of component $i$. The ratio $p_{v,i} / p_{sat,i}$ is the relative saturation of component $i$ (relative humidity in the case of water).
Here, the gas phase composition is generated by flowing nitrogen gas through a liquid bubbler containing an IPA-water mixture.
The gas phase composition should thus reflect the liquid-vapor equilibrium, up to an undersaturation that depends on the bubbler temperature.
This is tested in Figure~\ref{SIfig:Activity_RH_IPA-water}\textbf{b} by comparing the humidity values measured in the chamber as a function of the liquid IPA mass fraction in the bubbler with literature relative humidity values in carefully designed saturation conditions~\cite{brunjes1943vapor}.
We observe that the measured humidity values follow well the expected values but at a relative saturation of $\sim 80\%$.

\end{document}